\documentclass[preprint,12pt]{elsarticle}

\usepackage{amssymb}
\usepackage{amsmath}
\usepackage{float}
\usepackage{multirow}
\usepackage{algorithm}
\usepackage{algpseudocode}
\usepackage{gensymb}
\usepackage[hidelinks]{hyperref}
\usepackage{listings}
\usepackage{subcaption}
\usepackage{adjustbox}
\usepackage[table,dvipsnames,svgnames]{xcolor}      

\usepackage{caption}
\usepackage{graphicx}
\usepackage{latexsym}
\usepackage{times}
\usepackage[pagewise]{lineno}

\usepackage{fmtcount}
\usepackage{marvosym}
\usepackage{chemformula}
\usepackage{adjustbox}
\usepackage{epstopdf}

\usepackage{makecell}
\usepackage{nicematrix}
\usepackage{tikz}
\usepackage{dsfont}
\usetikzlibrary{spy}
\usepackage{siunitx}

\usepackage{bm}

\journal{}

\newcommand{\etal}{\emph{et al.}}
\newcommand{\pca}{PCA}
\newcommand{\cokpca}{CoK-PCA}
\newcommand{\sode}{sODE}
\newcommand{\node}{nODE}

\newcommand{\numg}{n_g}
\newcommand{\numv}{n_v}
\newcommand{\numq}{n_q}
\newcommand{\etamod}{\Tilde{\eta}}

\begin{document}
\sloppy
\begin{frontmatter}



\title{An {\em a-posteriori} analysis of co-kurtosis PCA based dimensionality reduction using a neural ODE solver}


\author[inst1]{Tadikonda Shiva Sai}
\author[inst2]{Hemanth Kolla}
\author[inst1]{Konduri Aditya\corref{cor1}}

\affiliation[inst1]{organization={Department of Computational and Data Sciences, Indian Institute of Science},
            city={Bengaluru},
            postcode={560012},
            state={KA},
            country={India}}

\affiliation[inst2]{organization={Sandia National Laboratories},
            city={Livermore},
            postcode={94550},
            state={CA},
            country={United States of America}}

\cortext[cor1]{Corresponding author.}



\begin{abstract}
A low-dimensional representation of thermochemical scalars based on cokurtosis principal component analysis (CoK-PCA) has been shown to effectively capture stiff chemical dynamics in reacting flows relative to the widely used principal component analysis (PCA). The effectiveness of the reduced manifold was evaluated in \textit{a priori} analyses using both linear and nonlinear reconstructions of thermochemical scalars from aggressively truncated principal components (PCs). In this study, we demonstrate the efficacy of a CoK-PCA-based reduced manifold using \textit{a posteriori} analysis. Simulations of spontaneous ignition in a homogeneous reactor that pose a challenge in accurately capturing the ignition delay time as well as the scalar profiles within the reaction zone are considered.  The governing ordinary differential equations (ODEs) in the PC space were evolved from the initial conditions using two ODE solvers. First, a standard ODE solver that uses a pre-trained artificial neural network (ANN) to estimate the source terms and integrates the solution in time. Second, a neural ODE solver that incorporates the time integration of PCs into the ANN training. The time-evolved profiles of the PCs and reconstructed thermochemical scalars demonstrate the robustness of the CoK-PCA-based low-dimensional manifold in accurately capturing the ignition process. Furthermore, we observed that the neural ODE solver minimized propagation errors across time steps and provided more accurate results than the standard ODE solver. The results of this study demonstrate the potential of CoK-PCA-based manifolds to be implemented in massively parallel reacting flow solvers.
\end{abstract}


\begin{keyword}

Cokurtosis-PCA; PCA; Dimensionality reduction; Neural ODE; Autoignition
\end{keyword}

\end{frontmatter}


\section{Introduction\label{sec:introduction}} \addvspace{10pt}
 High-fidelity combustion simulations, including direct numerical and large-eddy simulations, are often expensive because of the large number of computations that arise from the use of detailed chemical kinetics. The computational cost further increases for higher hydrocarbon fuels and with problems involving complex geometries. With statistical and machine learning techniques, evolving the kinetics on a low-dimensional manifold has drawn wide interest in significantly reducing chemistry costs. Among the different data-driven techniques for dimensionality reduction, the use of principal component analysis (\pca) and its variants has been widely investigated \cite{sutherland2009,biglari2012,mirgolbabaei2013,coussement2016,dalakoti2021,zdybal2023}. {\pca} based low-dimensional manifolds have been also successfully incorporated into scalable reacting flow solvers.

For instance, Echekki~\etal~\cite{echekki2015principal} performed simulations to analyze the transport of principal components (PCs) obtained from PCA, and demonstrated that the reconstructed thermochemical scalars were in excellent agreement with those obtained from the transport of the full thermochemical scalars. Biglari~\etal~\cite{biglari2015posteriori} employed PCA to model non-premixed temporally evolving jet flames with extinction and reignition. The performance of the model was examined by varying the number of retained principal components and its applicability across a range of Reynolds numbers was investigated. Recently, Kumar~\etal~\cite{kumar2023acceleration} performed 2D direct numerical simulations (DNS) of lean premixed methane-air flames stabilized on a slot burner and modelled the closure terms for PC transport equations. The surrogate models obtained from these simulations were used to perform 3D PC-DNS and validate the results against computationally expensive 3D species-DNS. {Malik~\etal~\cite{malik2021combustion} demonstrated the potential of {\pca} coupled with Gaussian process regression for three-dimensional large-eddy simulations of Sandia flames D, E, and F using only two principal components. In a recent study, Armstrong~\etal~\cite{armstrong2024reduced} employed a quantity of interest (QoI) based approach, where a linear projection was learned based on the reconstruction errors in the QoIs, such as source terms and state variables, and extensively studied the projected shape of the manifold along with the behavior of transport variables. Other studies that have demonstrated the applicability of low-dimensional manifolds in accurately simulating reacting flows include \cite{malik2024combined,abdelwahid2023}.

In general, {\pca} identifies the direction with the largest variance in the original data. Therefore, {\pca} may not effectively capture extremely valued samples, which are the characteristics of local spatio-temporal events, such as the formation of ignition kernels. Higher order moments have been shown to be better for capturing the statistical signatures of such events \cite{aditya2019anomaly}. To this end, the use of fourth-order joint statistical moments for dimensionality reduction, known as cokurtosis-based PCA (\cokpca), has been proposed \cite{jonnalagadda2023co}. \textit{A priori} analyses showed that low-dimensional manifolds based on {\cokpca} provides a more accurate representation of various species than {\pca} \cite{jonnalagadda2023co, nayak2024co}. Additionally, {\cokpca} was found to be more effective in predicting the heat release rate, indicating that it is better at capturing chemical kinetics. These studies demonstrated the efficacy of {\cokpca} in an \textit{a priori} setting by reconstructing thermochemical scalars from reduced PCs.  

In this study, we assess the effectiveness of {\cokpca} in modelling fuel oxidation in a homogeneous reactor using \textit{a posteriori} analysis, where the evolved thermochemical scalars from an initial condition are obtained by solving the governing ordinary differential equations (ODEs) in reduced PC space and nonlinear reconstruction. Additionally, we employ a neural ODE approach to model integrate the differential equations and investigate its efficacy. The remainder of this paper is organized as follows. Section~2 outlines the low-dimensional manifolds {\pca} and {\cokpca}. The procedure for integrating the ODEs is described in Sec.~3. Section 4 presents and analyzes the results and Sec.~5 presents the conclusions.}

\section{Low-dimensional manifolds\label{sec:ldms}} \addvspace{10pt}
A low-dimensional manifold for thermochemical scalars is generated using training data comprising trajectories obtained from different initial conditions that typically span the state space to be encountered during inference. Let the training data matrix be $\mathbf{D}\in R^{\numv\times \numg}$, where $\numg$ is the number of observations and $\numv$ is the number of variables or features (the mass fractions of different chemical species and the temperature). As the values in the matrix range several decades across different variables, scaling is commonly employed over each variable. Let $\mathbf{X}_{\numv \times \numg}$ be the scaled data matrix, where $x_i \in R^{1 \times \numg}$ denotes $i^{th}$ row of $\mathbf{X}$ and $x^{(i)} \in R^{\numv \times 1}$ denotes a single observation of state, that is, the $i^{th}$ column vector of $\mathbf{X}$. A low-dimensional representation of the data is then obtained by $\mathbf{Z}_q= \mathbf{A}_q \mathbf{X}$, where the low-dimensional manifold comprises $\numq$ variables, denoted by $\eta_i$ ($i=1,\dots,\numq$), such that $\numq < \numv$ and the linear transformation matrix is $\mathbf{A}_q \in R^{\numq\times \numv}$. In general, several choices exist for constructing matrix $\mathbf{A}_q$. As mentioned earlier, we consider two choices namely the principal component analysis (\pca) and the cokurtosis based principal component analysis (\cokpca), which will be detailed next.

\subsection{Principal component analysis (\pca)
\label{subsec:pca}} \addvspace{10pt}
Principal component analysis (\pca) is a commonly employed method for projecting high-dimensional data to lower dimensions by maximizing the variance captured within the data. The core idea is to identify an orthogonal basis for the data space ordered by the variance explained along each basis vector. These basis vectors, known as principal vectors, are used to project the data, resulting in principal components (PCs). The first principal vector aligns with the direction of maximum variance in the data, and each subsequent vector corresponds to the direction of the maximum variance within the subspace orthogonal to the preceding vectors. The computation of these vectors requires the data covariance matrix $\mathbf{C}$ which is defined as  
\begin{equation}  
\left(\mathbf{C}\right)_{ij} \equiv C_{ij} = \mathds{E}\left(x_i x_j\right) \quad i, j \in \{1, 2, ..., n_v\}, 
\label{eq:covariance}  
\end{equation} 
where $\mathds{E}$ denotes the expectation operator, and $x_i$ represents the original features, centered around their respective means. The required principal vectors, which form the rows of projection matrix $\mathbf{A}$, are the eigenvectors of the covariance matrix. These are determined through the eigenvalue decomposition of the covariance matrix, expressed as $\mathbf{C}=\mathbf{Q}\mathbf{\Lambda} \mathbf{Q}^{T}$, where $\mathbf{A}=\mathbf{Q}^{T}$.  The eigenvalues (diagonal elements of $\mathbf{\Lambda}$) indicate the variances captured along the corresponding principal vectors.

\subsection{Co-kurtosis based principal component analysis (\cokpca)
\label{subsec:cokpca}} \addvspace{10pt}
For combustion datasets, it has been argued that variance maximization is not necessarily the best choice. Recently, a change in basis that optimizes for a higher-order moment, particularly cokurtosis, has been proposed \cite{jonnalagadda2023co}. The central principle is that the stiff dynamics of chemically reacting systems are better represented by regions in the state space that correspond to outlying samples, and these portions of the state space are better represented by directions representing higher-order statistical moments, such as kurtosis, rather than variance \cite{aditya2019anomaly}. Mathematically, the cokurtosis-based {\pca} (\cokpca) technique considers the fourth-order cumulant tensor $\mathbf{T}$: 
\begin{equation} 
T_{ijkl} = \mathds{E}(x_i x_j x_k x_l) - C_{ij} C_{kl} - C_{ik} C_{jl} - C_{il} C_{jk}. 
\label{eq:coktensor}
\end{equation}
The basis vectors are identified by factorizing $\mathbf{T}$, which is a symmetric fourth-order tensor. Among the various methods for factorizing symmetric higher-order tensors, simple higher-order singular value decomposition (HOSVD) has been shown to be robust and useful due to the orthogonality imposed on the resulting vectors \cite{aditya2019anomaly}. The tensor $\mathbf{T}$ is reshaped into $n_v \times n_v^3$ matrix, and the columns of the left singular matrix $\mathbf{U}$ obtained from the SVD of $matricized(\mathbf{T})=\mathbf{U}\mathbf{\Sigma} \mathbf{V}^T$ provide the required principal vectors ($\mathbf{A} = \mathbf{U}^{T})$.

\section{Solver details \label{sec:odesolvers}} \addvspace{10pt}

\subsection{Governing equations
\label{subsec:ge}} \addvspace{10pt}
For the homogeneous reactor that captures the spontaneous ignition of fuel-oxidizer mixtures, the governing equations to evaluate the thermochemical scalars are 
\begin{equation}    
\frac{d\boldsymbol{\Phi}}{dt} = \boldsymbol{S_{\Phi}},     \label{eq:eqn_phi} 
\end{equation}
where $\boldsymbol{\Phi} = [Y_1, Y_2,\dots,Y_{n_s-1}, T]^T \in R^{n_s\times 1}$ is a vector of species mass fractions and temperature, $\boldsymbol{S_\Phi} \in R^{n_s\times 1}$ is a vector of the corresponding chemical source terms, and $n_s$ is the number of species considered in detailed chemistry. The columns of data matrix $\mathbf{D}$ correspond to $\boldsymbol{\Phi}$ at different time steps. Using the projection matrix $\boldsymbol{A}$, data in principal component (PC) space can be obtained as {$\boldsymbol{\eta}=\boldsymbol{A}\boldsymbol{x}$, where $\boldsymbol{x}$ is the scaled vector corresponding to $\boldsymbol{\Phi}$}. The equation evolving the PCs is  \begin{equation}     \frac{d\boldsymbol{\eta}}{dt} = \boldsymbol{S_{\eta}},     \label{eq:eqn_eta} \end{equation} with initial condition $\boldsymbol{\eta}(0) = \boldsymbol{A} (\boldsymbol{x}(0))$. The source term for the PCs can be obtained as $\boldsymbol{S_{\eta}} = \boldsymbol{A} \boldsymbol{S_x}$, where $\boldsymbol{S_x}$ is a scaled vector corresponding to $\boldsymbol{S_\Phi}$. The above equation is a linear transformation of Eq.~\ref{eq:eqn_phi}. For $\numv$ thermochemical scalars, both {\pca} and {\cokpca} provide $\numv$ PCs. A low-dimensional manifold is obtained by considering only a subset of PCs $\numq$, and the truncated transformation matrix is $\boldsymbol{A_q}$. To reconstruct thermochemical scalars from PCs \cite{jonnalagadda2023co}, the naive approach is to use the inverse of the eigenvector matrix ($\mathbf{A}^{-1}$). However, because the linear transformation from the full thermochemical state space to the reduced state space is not invertible, a simple linear transformation cannot accurately reconstruct thermochemical scalars. Therefore, nonlinear reconstruction methods such as ANNs are employed, which will be used in this study. Next, we describe the procedure for evolving the PCs, which includes the ANN-based modelling of the source terms. 

\subsection{Standard ODE (\sode) solver
\label{subsec:sode}} 
\begin{figure}[h!]
    \centering
    \includegraphics[width=13cm]{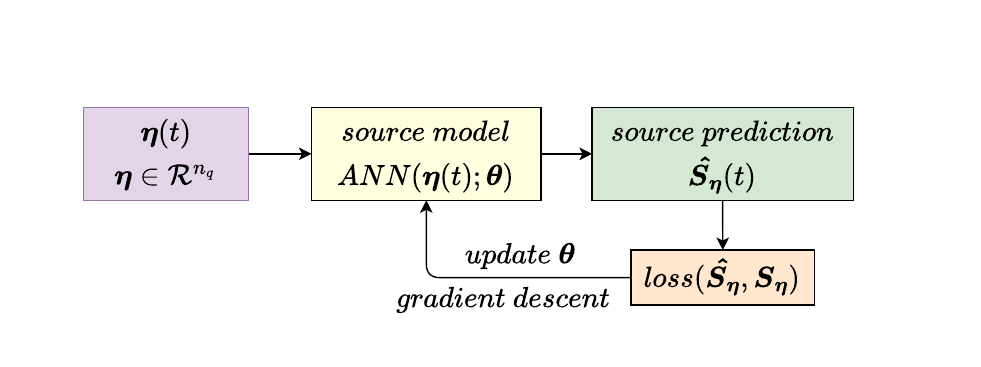}
    \caption{\footnotesize Schematic of ANN training to predict the source terms ($\boldsymbol{S_{\eta}}$) in the {\sode} solver.}
    \label{fig:sODE}
\end{figure}

\addvspace{10pt}
The integration of the governing equations (Eq.~\ref{eq:eqn_eta}) in the low-dimensional space requires the source terms $\mathbf{S_{\eta}}$ at each time step. However, there is no direct method for computing these source terms solely as a function of the PCs. Therefore, pre-trained artificial neural networks (ANNs) are employed to predict source terms during time integration \cite{echekki2015principal, owoyele2017toward, kumar2023acceleration}. The implementation of this approach is referred to as the standard ODE (\sode) solver. Figure~\ref{fig:sODE} shows the standard training framework used to optimize the network parameters for learning source terms. \textit{A priori} PC profiles are used as inputs to the network, while the projected source terms ($\boldsymbol{S_{\eta}} = \boldsymbol{A} \boldsymbol{S_x}$) are used to compute the training loss. Once trained, the network can be combined with conventional numerical schemes to evolve the PCs. Furthermore, the corresponding thermochemical scalars ($\boldsymbol{\Phi}$) are obtained by using nonlinear reconstruction, as mentioned earlier.

 The {\sode} solver introduces three primary sources of error: (1) approximation errors in the source terms of the PCs, (2) errors from the numerical scheme used to solve the governing ODEs, and (3) reconstruction errors for thermochemical scalars. In the context of dimensionality reduction, errors from source-term approximations are critical to obtain an accurate solution. Although the source-term network can be trained with a greater accuracy, there will invariably be small errors in the source-term predictions for some observations in the dataset. It is essential to ensure that these errors which would propagate in time, do not amplify and result in an unstable solution.
 

\subsection{Neural ODE (\node) solver
\label{subsec:node}} 
\begin{figure}[h!]
    \centering
    \includegraphics[width=13cm]{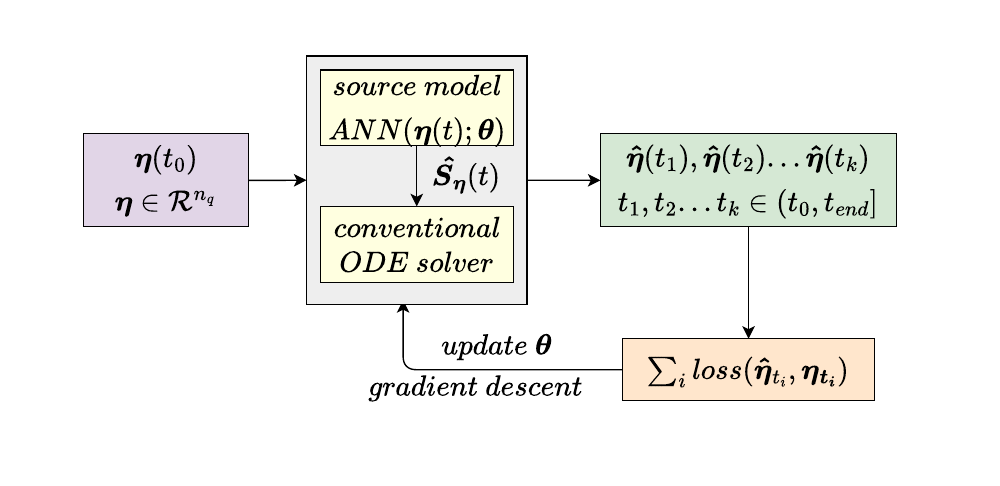}
    \caption{\footnotesize Schematic of ANN training to predict the source terms ($\boldsymbol{S_{\eta}}$) in the {\node} solver.}
    \label{fig:nODE}
\end{figure}

\addvspace{10pt}
 To ensure that the approximation errors in the source terms at each time step are not amplified in time, the time advancement of PCs can be incorporated into the source-term model training, which can be achieved using the neural ODE framework \cite{10.5555/3327757.3327764, dikeman2022stiffness, kim2021stiff, owoyele2022chemnode}. Neural ODEs enable source-term learning by minimizing the errors in the evolved PCs, thereby ensuring the stability of the solution during the training phase. The neural ODE (\node) solver comprises two primary components: an artificial neural network for modelling the source terms, and a conventional numerical solver for evolving the ODE system. Given an initial condition as input, the model computes the solution at the desired time steps. The loss is computed directly for the obtained PCs, and the weights of the network modelling the source term are optimized to minimize these errors. A schematic of the training process is shown in Fig.~\ref{fig:nODE}.

The primary challenge in training {\node}s is the computation of gradients with respect to the parameters of the network, which traditionally involves backpropagation through all operations of the ODE solver. This requires storing all intermediate outputs from the forward pass, making it memory-intensive, and introducing additional numerical errors. To address this issue, Chen \etal~\cite{10.5555/3327757.3327764} proposed a more efficient method for computing gradients by numerically integrating an augmented ODE system backwards in time. This approach can be viewed as a continuous analogue to discrete reverse-mode automatic differentiation.

In certain cases, we observe that the ANN in the {\node} solver converges to a suboptimal local minimum. In such scenarios, we propose a small modification; instead of learning the evolution of PCs ($\boldsymbol{\eta}(t)$), the focus is on learning $\etamod_i(t)=\eta_i(t)-\eta_i(t_0)$, where $\eta_i(t_0)$ is the initial condition. Accordingly, the governing equation can be rewritten as 
\begin{equation}       
\frac{d\boldsymbol{\etamod}}{dt} = \boldsymbol{S_{\etamod}},       \label{eq:eqn_eta_modified} 
\end{equation}
where source term $\boldsymbol{S_{\etamod}}(t)$ is identical to $\boldsymbol{S_{\eta}}(t)$ and is now a function of both $\boldsymbol{\etamod}(t)$ and $\boldsymbol{\eta}(t_0)$. During the implementation, $\boldsymbol{\etamod}(t)$ and $\boldsymbol{\eta}(t_0)$ are concatenated and provided as inputs to the ANN to obtain $\boldsymbol{{S_{\etamod}}}$. Using this approach, we found that the model converges to a better local minimum. A plausible explanation is that in the earlier approach, during training, if the PCs predictions from different configurations converge or coincide at any time step, the solution at later steps would be identical, regardless of the initial condition. This phenomenon may be the underlying cause of convergence to suboptimal local minima. With the proposed modification, the initial condition is consistently provided as an input to the network. Consequently, even if $\boldsymbol{\etamod}(t)$ coincides at any time step for two different configurations, the input at the subsequent step would remain distinct for both, potentially leading to a better local minimum.

\section{Results\label{sec:results}} \addvspace{10pt}

Fuel oxidation in a constant-pressure homogeneous reactor is used as a test case to investigate the performance of {\pca} and {\cokpca} in an \textit{a posteriori} setting. Specifically, we consider two configurations that use ethylene-air and nheptane-air as fuel-oxidizer mixtures to simulate the autoignition process. The reactor pressure is maintained at \SI{1.72}{atm}, with an initial temperature ($T$) ranging from \SI{1100}{\kelvin} to \SI{1300}{\kelvin} and an equivalence ratio ($\phi$) from \SI{0.5}{} to \SI{0.9}{}. The dataset for training and evaluation is obtained using \emph{Cantera} \cite{goodwin20238137090}, which evolves thermochemical scalars for various initial conditions.    

The {\sode} and {\node} solvers are implemented in \emph{Julia}  using the \texttt{DifferentialEquations.jl} and \texttt{Flux.jl} packages \cite{bezanson2017julia,rackauckas2017differentialequations,innes:2018}. For numerical integration, \textit{AutoTsit5} with the \textit{Rosenbrock23} scheme is employed. In the {\node} solver, \textit{BacksolveAdjoint} from the \texttt{SciMLSensitivity.jl} package \cite{rackauckas2020universal} is used for backpropagation. Prior to ANN training, the inputs and outputs of the network are normalized as a standard practice. PCs, which are part of the input in the {\sode} solver and both the input and output in the {\node} solver, are scaled to a zero mean and unit standard deviation. In the case of ANNs that are used to reconstruct thermochemical scalars, the input PCs are scaled to zero mean and unit standard deviation, whereas the output mass fractions and temperature are scaled to an interval $[0,1]$.

To ensure a fair comparison between {\pca} and {\cokpca}, hyperparameter tuning is performed on a subset of the training configurations using the \texttt{HyperTuning.jl} package \cite{mejia2021automated}. The ANN models are then trained on the full training dataset using optimal hyperparameters. The Adam optimizer is used in all cases. Gradient clipping is implemented in the tuned ANN models to mitigate instabilities caused by exploding gradients during optimization.  Despite this measure, some stiff regions were encountered in the parameter space, resulting in a significant increase in the training iteration time. In such scenarios, decreasing the learning rate by a factor of ten improved the training time. During the inference, PC values are obtained at uniformly spaced time steps. 

\subsection{Spontaneous ignition of ethylene-air mixture
\label{subsec:ethylene_results}} \addvspace{10pt}
The ethylene-air chemistry in the training and test datasets is represented by a 32-species, 206-reactions mechanism \cite{luo2012chemical}. {\pca} and {\cokpca} based low-dimensional manifolds (LDMs) are generated using thermochemical scalar data obtained from 13 distinct initial conditions. These data are also used to train the ANN models in {\sode} and {\node} solvers. Separately, data from ten randomly selected initial conditions within the training distribution are used for validation. To evaluate the trained models, test data consisting of ten additional configurations farthest from the training and validation sets are considered. The covariance matrix and cokurtosis tensor to generate the LDMs are computed using scaled data, where each feature is scaled using Pareto scaling. After obtaining the principal components (PCs), the dimensionality of the LDMs is set to $n_q=5$ \cite{nayak2024co}.

\begin{figure}[h!]
\newcommand{\figwidth}{1.8in}
\newcommand{\figheight}{1.1in}
\newcommand{\xlabelx}{-70}
\newcommand{\xlabely}{0}
\newcommand{\ylabelx}{-140}
\newcommand{\ylabely}{40}
\centering
\includegraphics[width=\figwidth, height=\figheight]{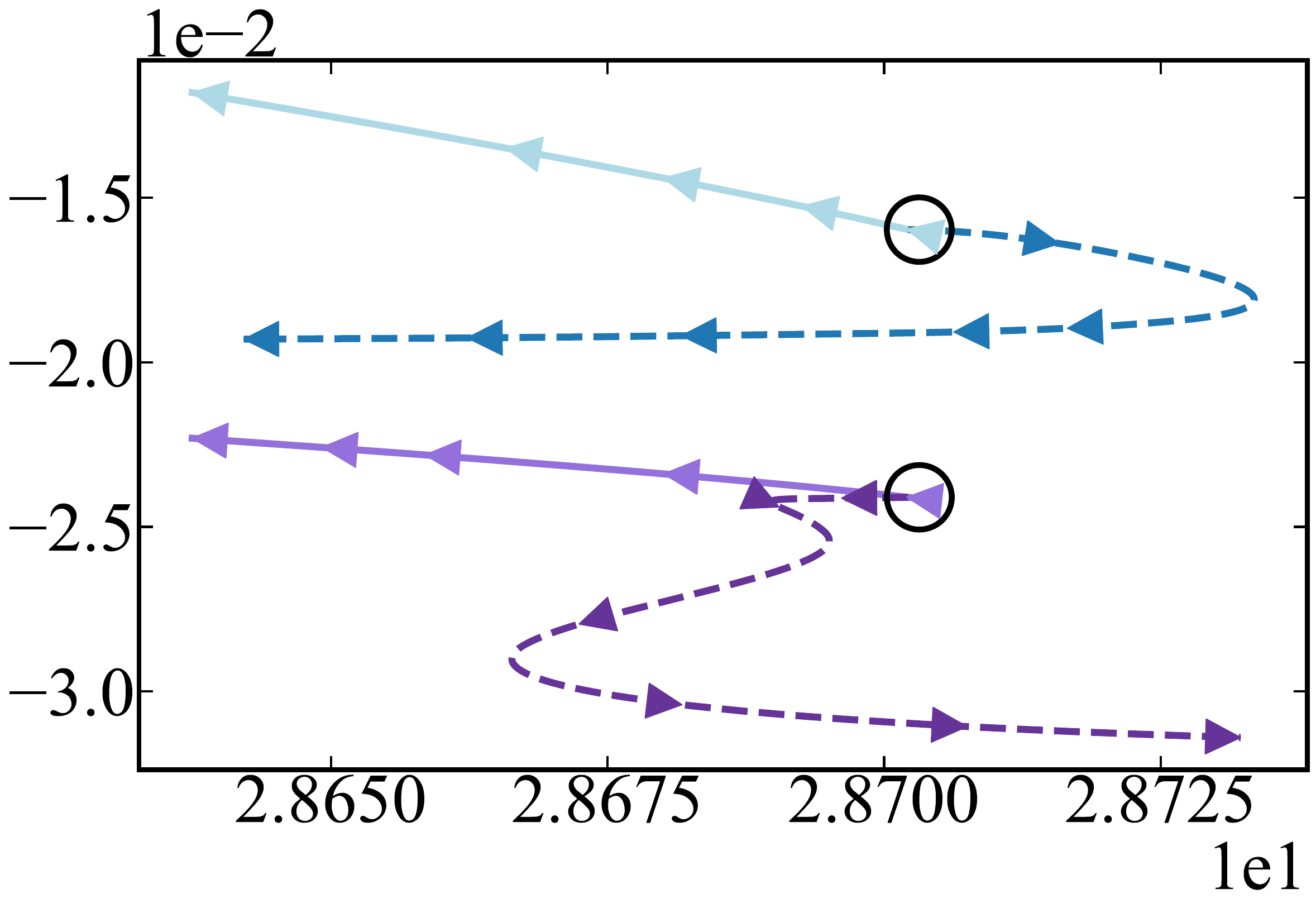}
\begin{picture}(0,0)
       \put(\xlabelx, \xlabely){\scriptsize $\eta_1$}
        \put(\ylabelx, \ylabely){\footnotesize{\rotatebox{90}{$\eta_3$}}}
    \end{picture}
\vspace{0.1cm}
\caption{\footnotesize Trajectories of the first and third PCs for one of the training configurations ($T=1200K, \phi = 0.5$) using {\sode} with {\pca} (blue) and {\cokpca} (purple). Solid and dashed lines represent the \textit{a priori} and \textit{a posteriori} solutions, respectively.}
\label{fig:sode_results}
\end{figure}
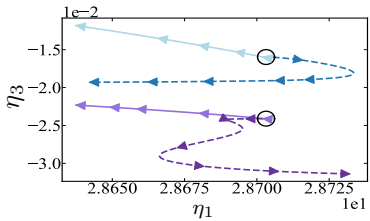
First, we assess the PC profiles computed using the {\sode} solver. Figure \ref{fig:sode_results} illustrates the time trajectories of the first and third PCs obtained from the solver for one of the training configurations, along with their \textit{a priori} counterparts. We observe that both the {\pca} and {\cokpca} profiles exhibit significant deviations from their \textit{a priori} trajectories, indicating divergent or unstable solutions. This is attributed to the accumulation of ANN model errors in time, which, even with small levels of inaccuracies, perturbs the thermal runaway of the ignition process, resulting in divergent subsequent trajectories. Note that the trajectories from {\cokpca} diverge slightly later than those from {\pca}. A similar trend is observed for the other configurations within the training data. Because the {\sode} solver results in divergent solutions, we do not consider this approach for further analysis.

\begin{figure*}[h!]
    \newcommand{\figwidth}{0.9in}
    \newcommand{\figheight}{0.8in}
    \newcommand{\xlabelx}{-40}
    \newcommand{\xlabely}{0}
    \newcommand{\ylabelx}{-74}
    \newcommand{\ylabely}{30}
    \newcommand{\figspacex}{0.05cm}
    \centering
    \includegraphics[width=\figwidth, height=\figheight]{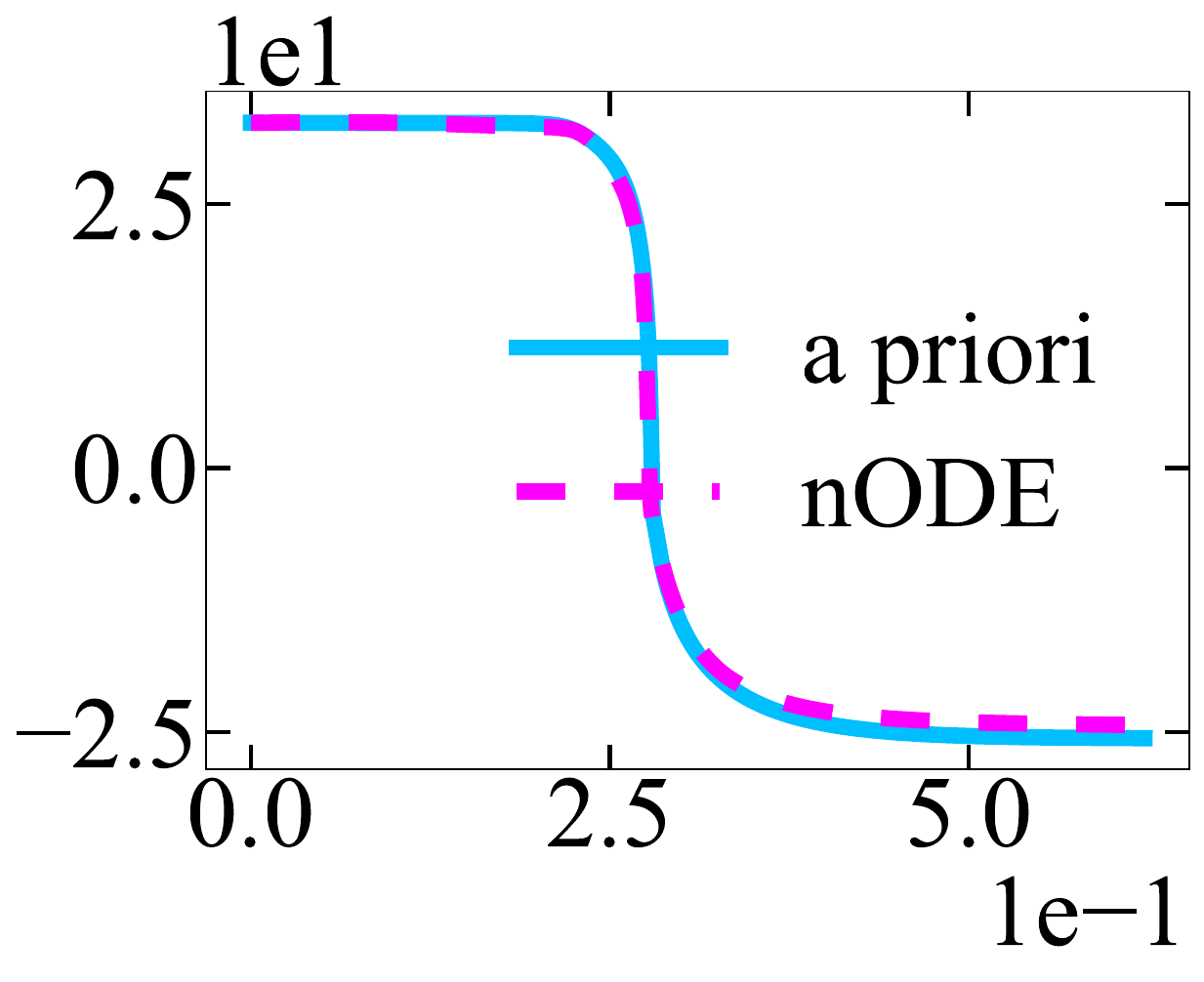}
    \begin{picture}(0,0)
       \put(\xlabelx, \xlabely){\scriptsize t (ms)}
        \put(\ylabelx, \ylabely){\footnotesize{\rotatebox{90}{$\eta_1$}}}
    \end{picture}
    \hspace{\figspacex}
    \includegraphics[width=\figwidth, height=\figheight]{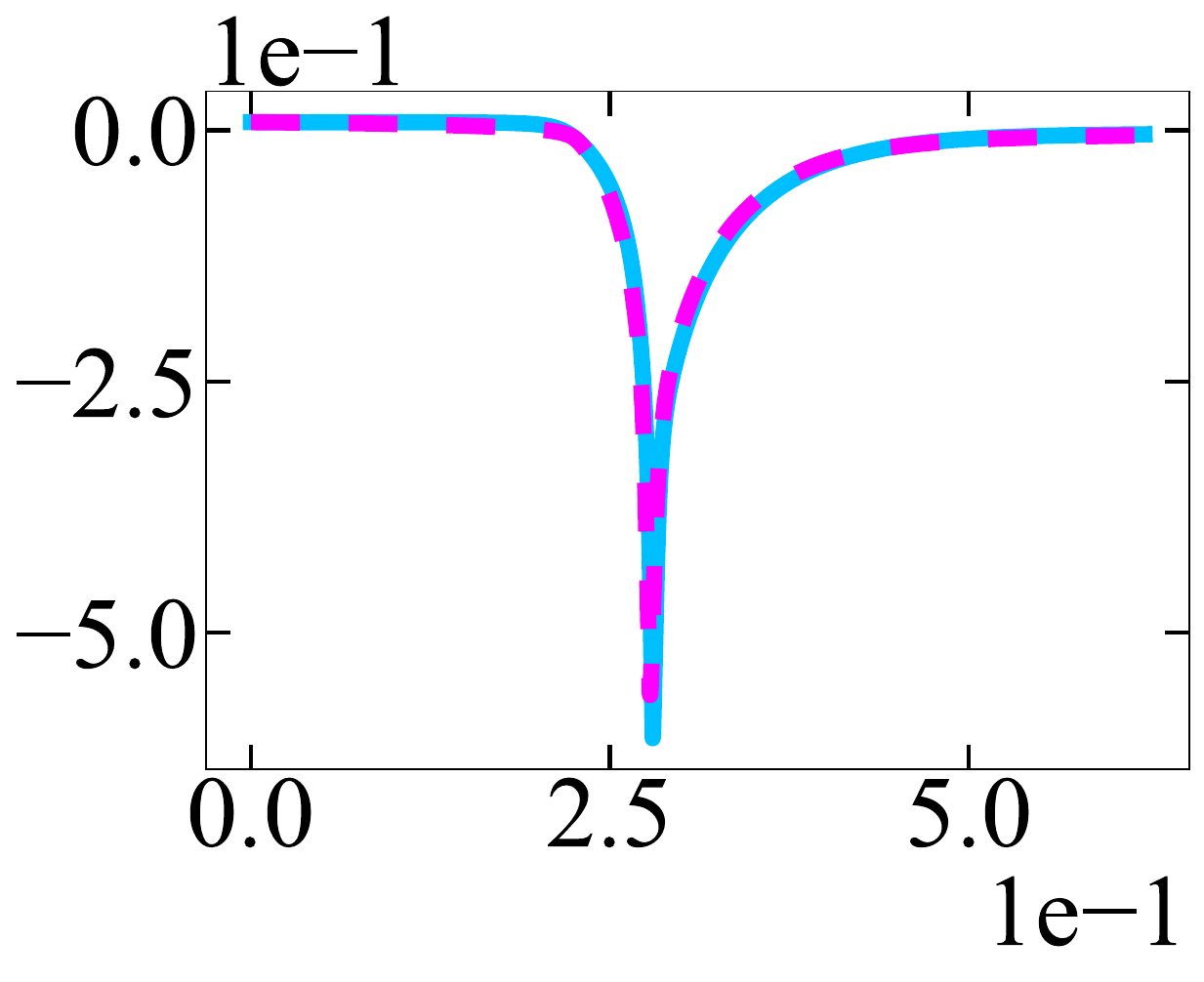}
    \begin{picture}(0,0)
       \put(\xlabelx, \xlabely){\scriptsize t (ms)}
        \put(\ylabelx, \ylabely){\footnotesize{\rotatebox{90}{$\eta_2$}}}
    \end{picture}
    \hspace{\figspacex}
    \includegraphics[width=\figwidth, height=\figheight]{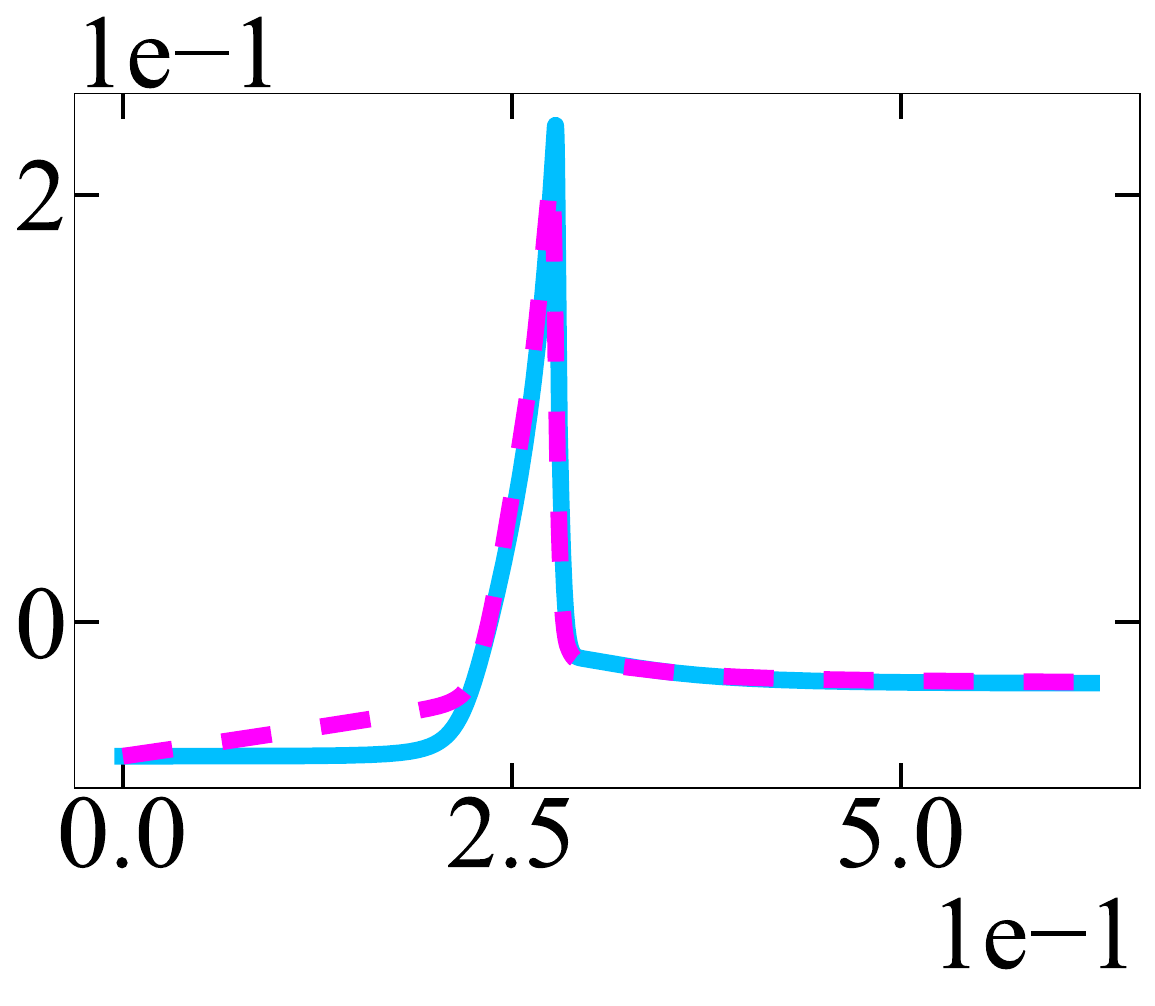}
    \begin{picture}(0,0)
       \put(\xlabelx, \xlabely){\scriptsize t (ms)}
        \put(\ylabelx, \ylabely){\footnotesize{\rotatebox{90}{$\eta_3$}}}
    \end{picture}
    \hspace{\figspacex}
    \includegraphics[width=\figwidth, height=\figheight]{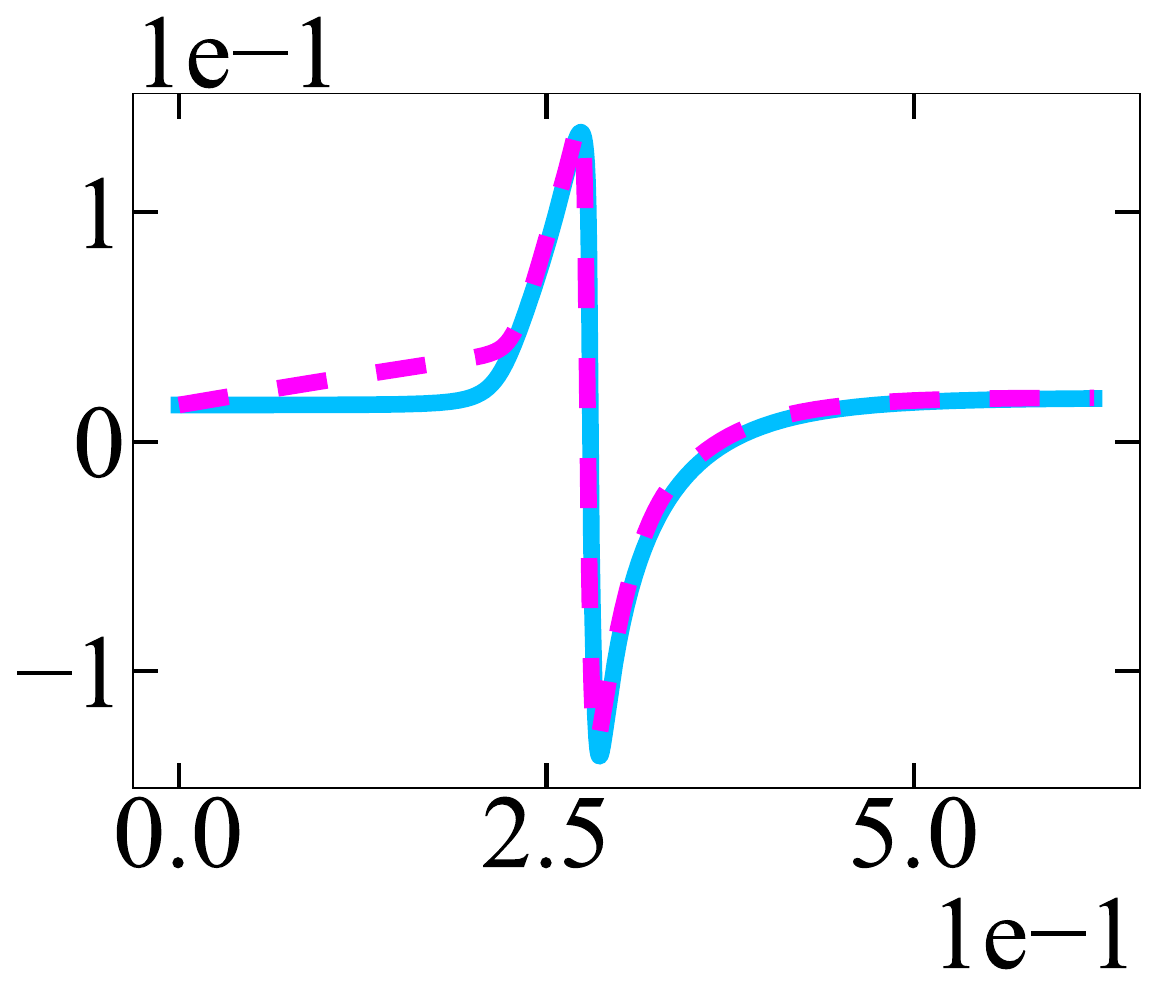}
    \begin{picture}(0,0)
       \put(\xlabelx, \xlabely){\scriptsize t (ms)}
        \put(\ylabelx, \ylabely){\footnotesize{\rotatebox{90}{$\eta_4$}}}
    \end{picture}
    \hspace{\figspacex}
    \includegraphics[width=\figwidth, height=\figheight]{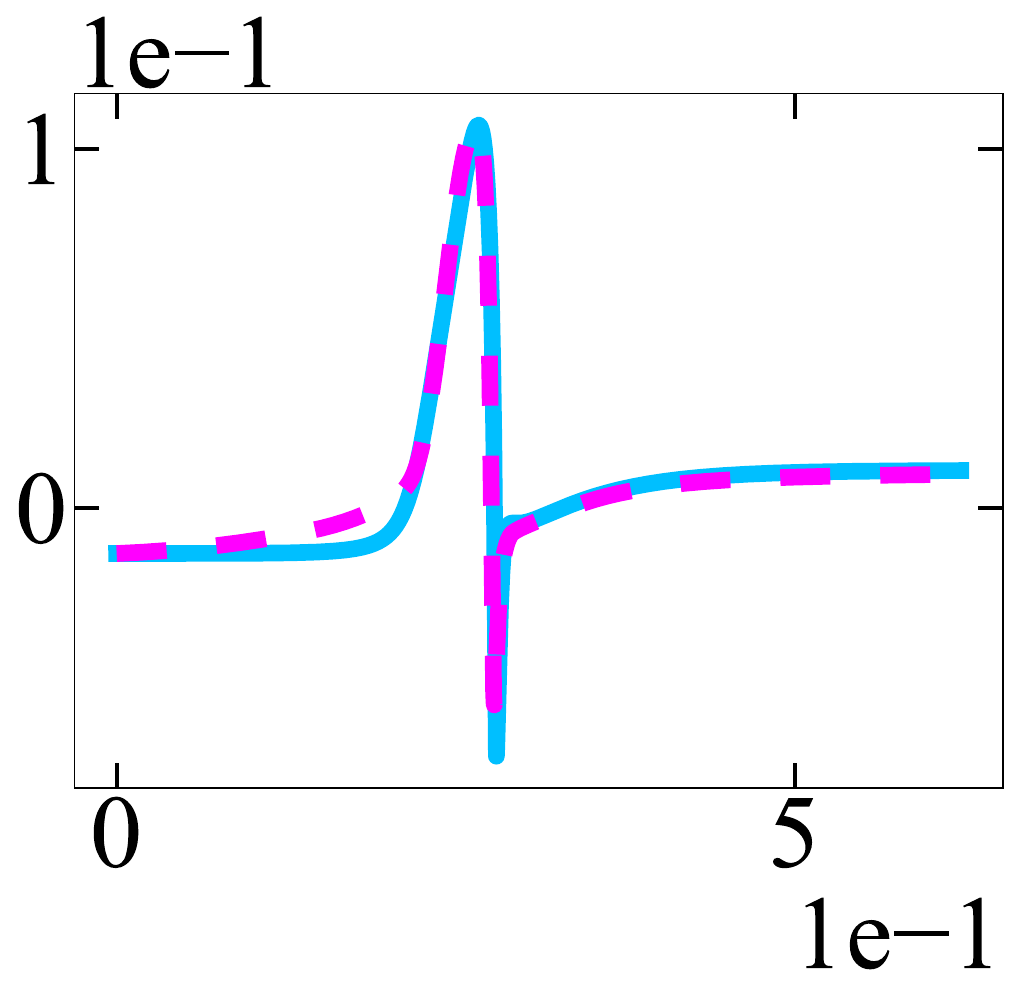}
    \begin{picture}(0,0)
       \put(\xlabelx, \xlabely){\scriptsize t (ms)}
        \put(\ylabelx, \ylabely){\footnotesize{\rotatebox{90}{$\eta_5$}}}
    \end{picture}
    
    \includegraphics[width=\figwidth, height=\figheight]{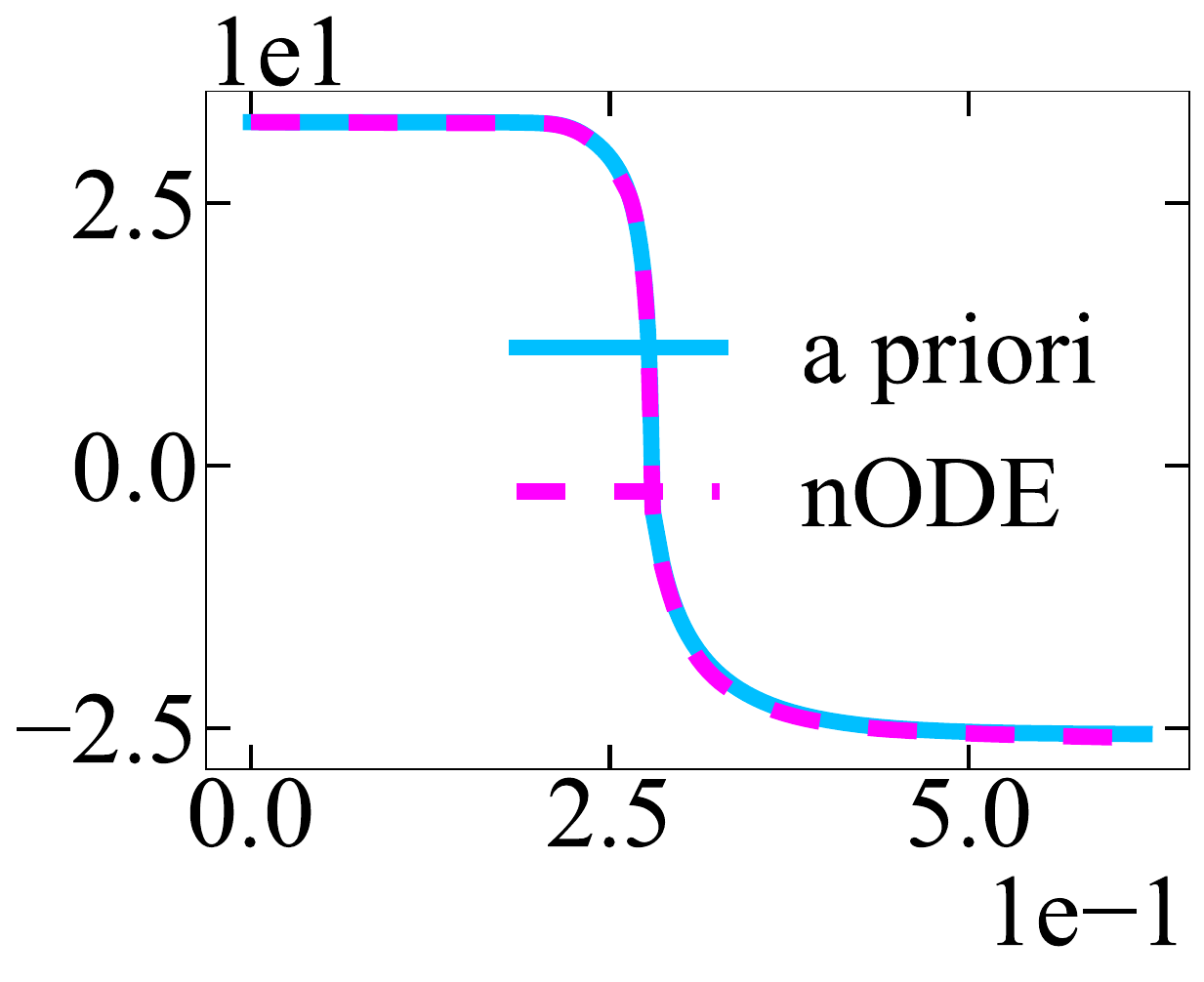}
    \begin{picture}(0,0)
       \put(\xlabelx, \xlabely){\scriptsize t (ms)}
        \put(\ylabelx, \ylabely){\footnotesize{\rotatebox{90}{$\eta_1$}}}
    \end{picture}
    \hspace{\figspacex}
    \includegraphics[width=\figwidth, height=\figheight]{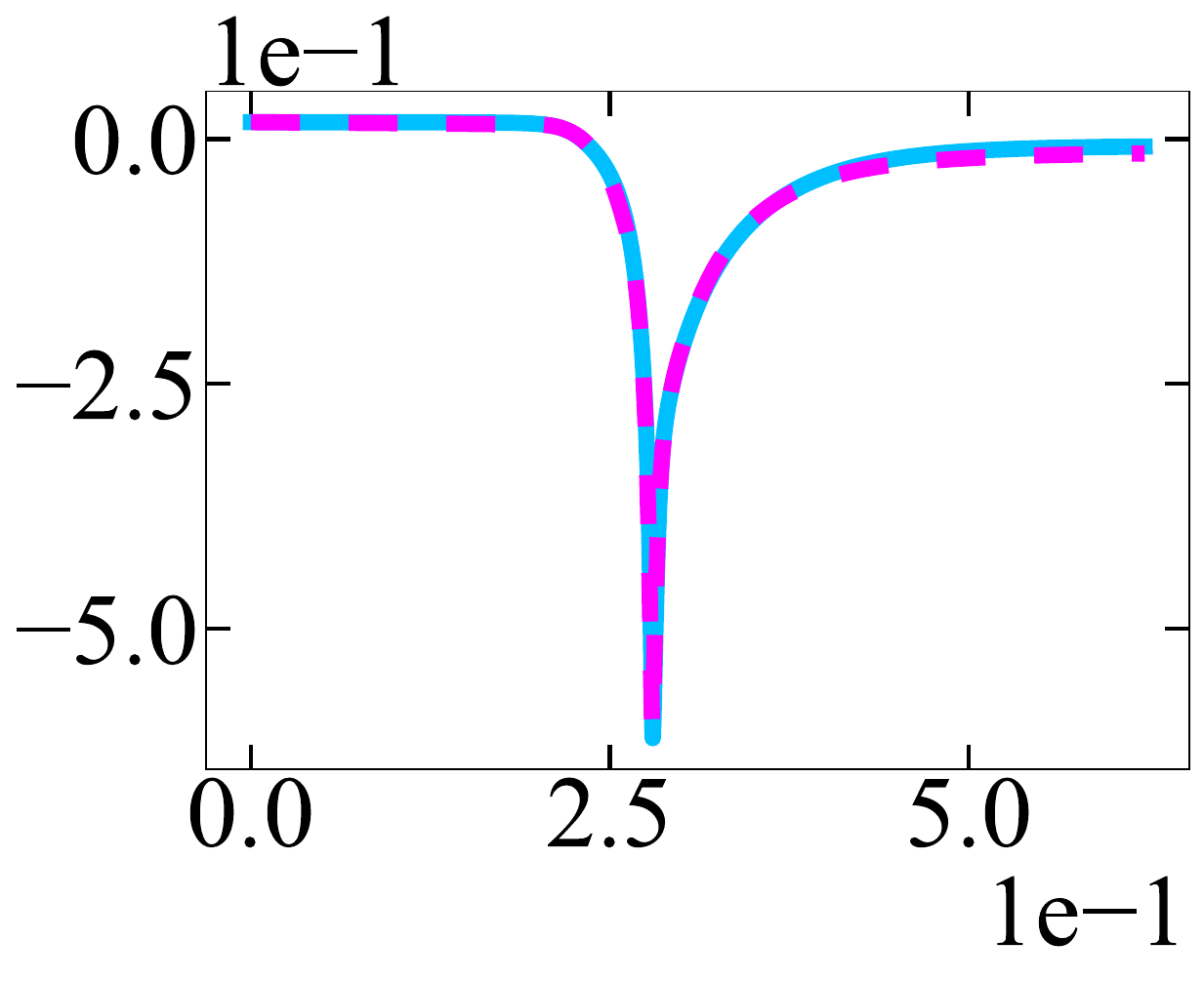}
    \begin{picture}(0,0)
       \put(\xlabelx, \xlabely){\scriptsize t (ms)}
        \put(\ylabelx, \ylabely){\footnotesize{\rotatebox{90}{$\eta_2$}}}
    \end{picture}
    \hspace{\figspacex}
    \includegraphics[width=\figwidth, height=\figheight]{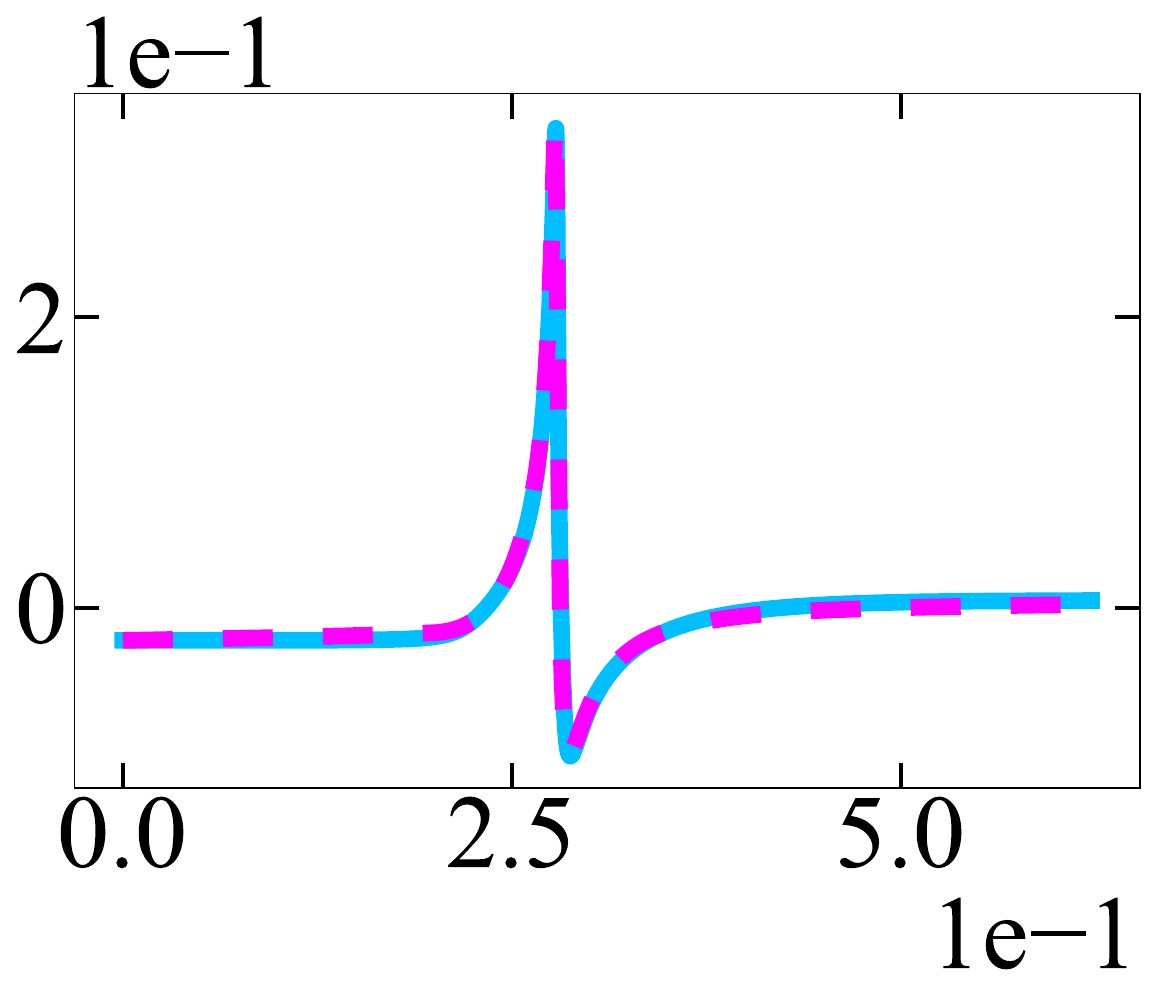}
    \begin{picture}(0,0)
       \put(\xlabelx, \xlabely){\scriptsize t (ms)}
        \put(\ylabelx, \ylabely){\footnotesize{\rotatebox{90}{$\eta_3$}}}
    \end{picture}
    \hspace{\figspacex}
    \includegraphics[width=\figwidth, height=\figheight]{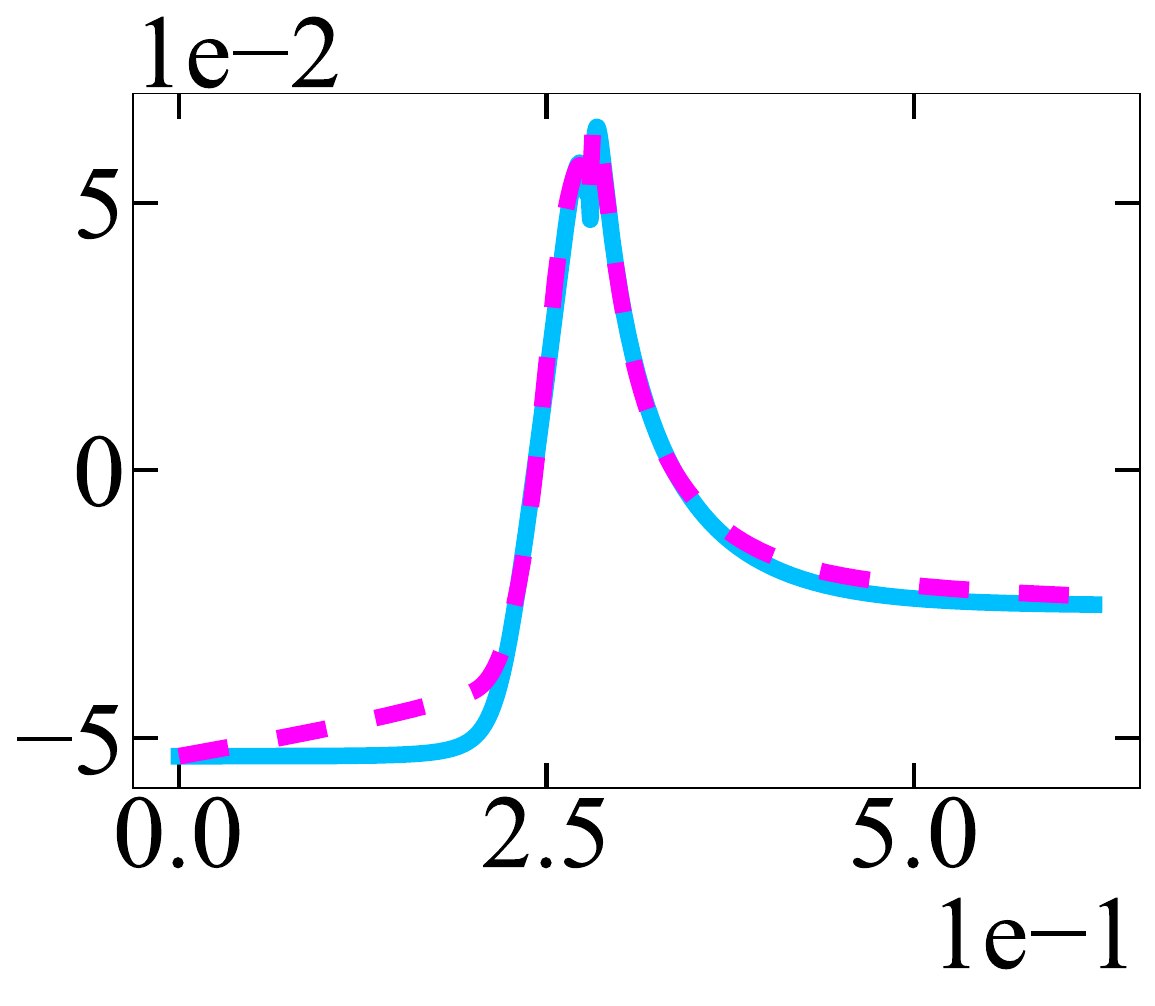}
    \begin{picture}(0,0)
       \put(\xlabelx, \xlabely){\scriptsize t (ms)}
        \put(\ylabelx, \ylabely){\footnotesize{\rotatebox{90}{$\eta_4$}}}
    \end{picture}
    \hspace{\figspacex}
    \includegraphics[width=\figwidth, height=\figheight]{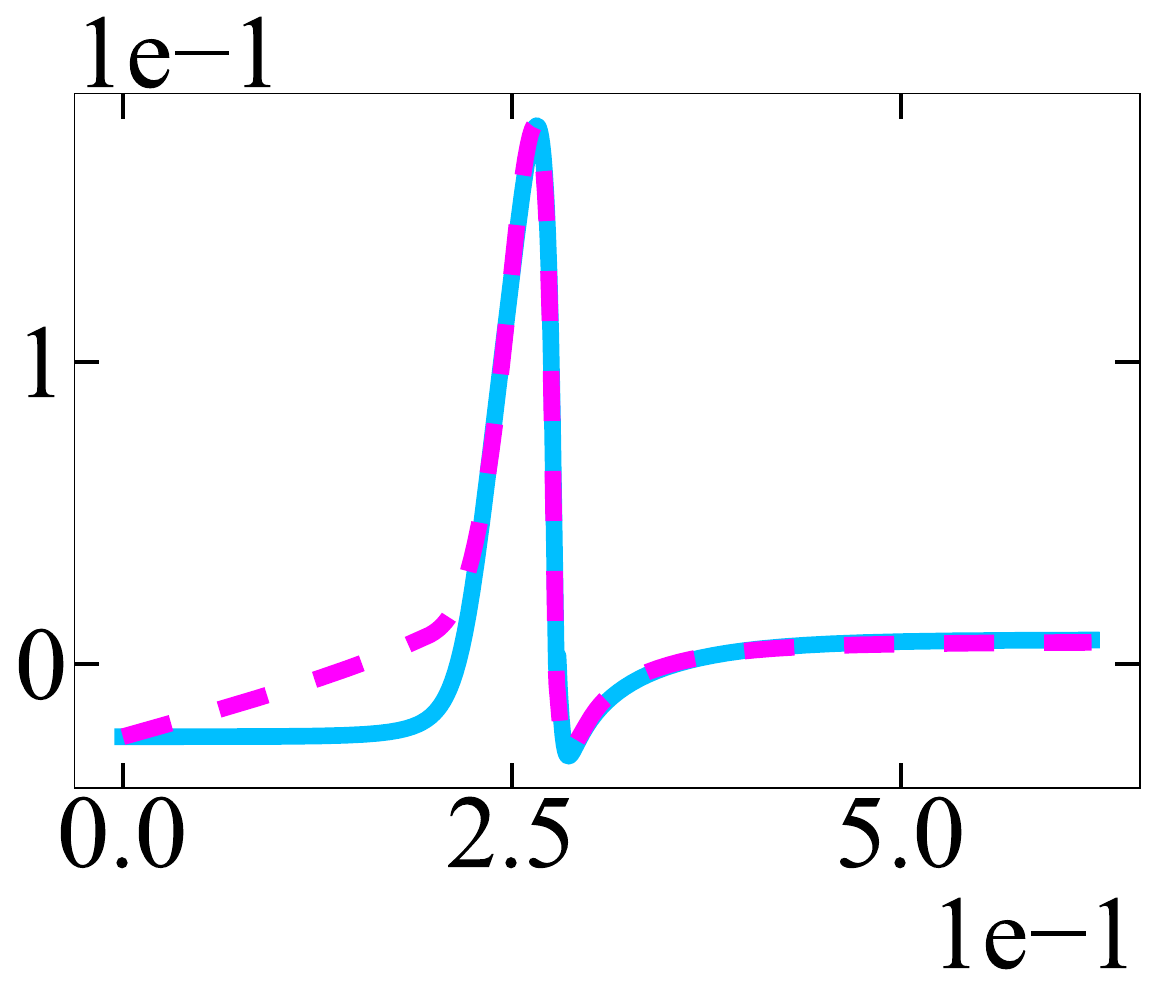}
    \begin{picture}(0,0)
       \put(\xlabelx, \xlabely){\scriptsize t (ms)}
        \put(\ylabelx, \ylabely){\footnotesize{\rotatebox{90}{$\eta_5$}}}
    \end{picture}

    \vspace{0.1cm}
    \caption{\footnotesize Temporal profiles of PCs from {\pca} (top row) and {\cokpca} (bottom row) based LDMs for a test configuration. A posteriori profiles are computing using the {\node} solver. A priori profiles are also shown for reference.}
    \label{fig:PCs_node_ethylene}
\end{figure*}
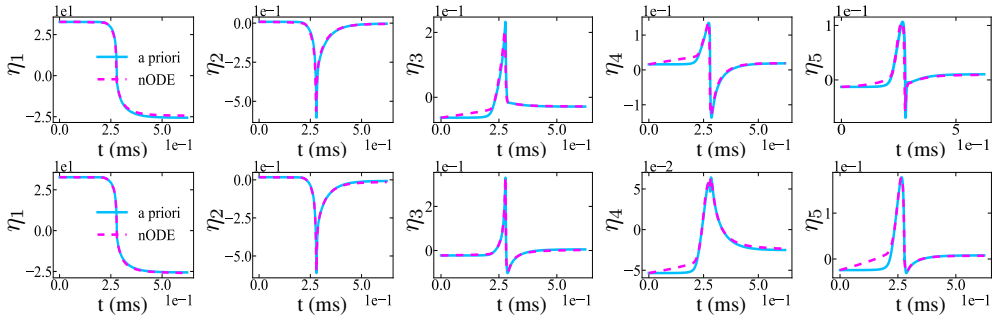

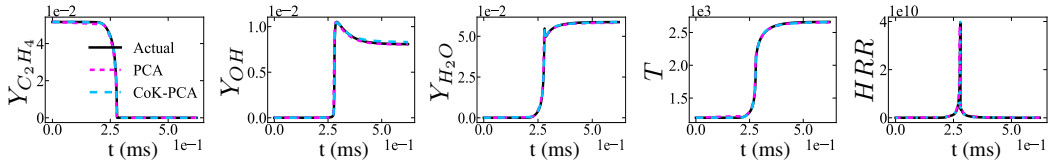
\begin{figure*}[h!]
\newcommand{\figwidth}{0.9in}
\newcommand{\figheight}{0.8in}
\newcommand{\xlabelx}{-40}
\newcommand{\xlabely}{0}
\newcommand{\ylabelx}{-73}
\newcommand{\ylabely}{32}
\newcommand{\figspacex}{0.2cm}
\centering
\includegraphics[width=\figwidth, height=\figheight]{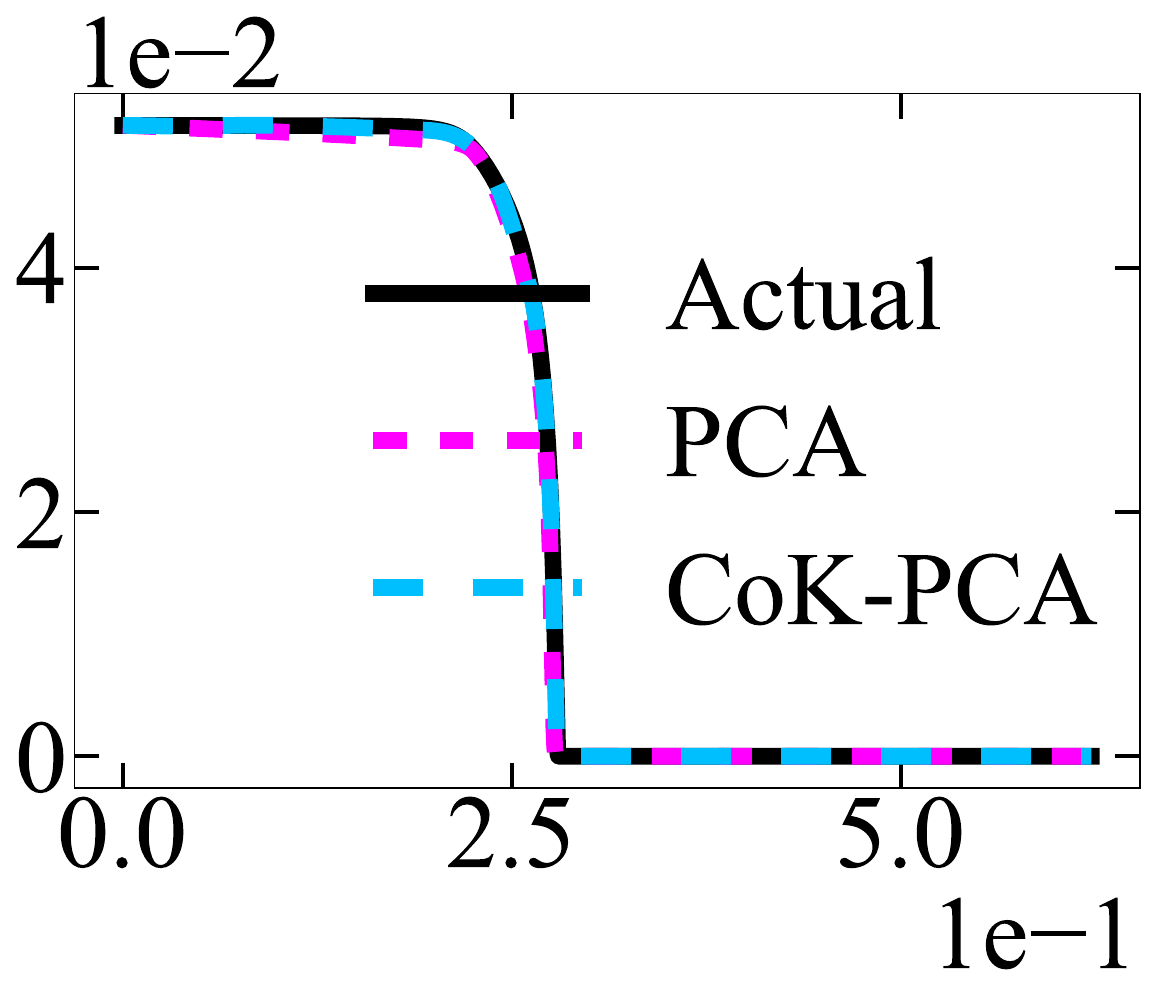}
\begin{picture}(0,0)
        \centering
       \put(\xlabelx, \xlabely){\scriptsize t (ms)}
        \put(\ylabelx, \ylabely){\makebox(0,0){\footnotesize{\rotatebox{90}{$Y_{C_2H_4}$}}}}
    \end{picture}
\hspace{\figspacex}
\includegraphics[width=\figwidth, height=\figheight]{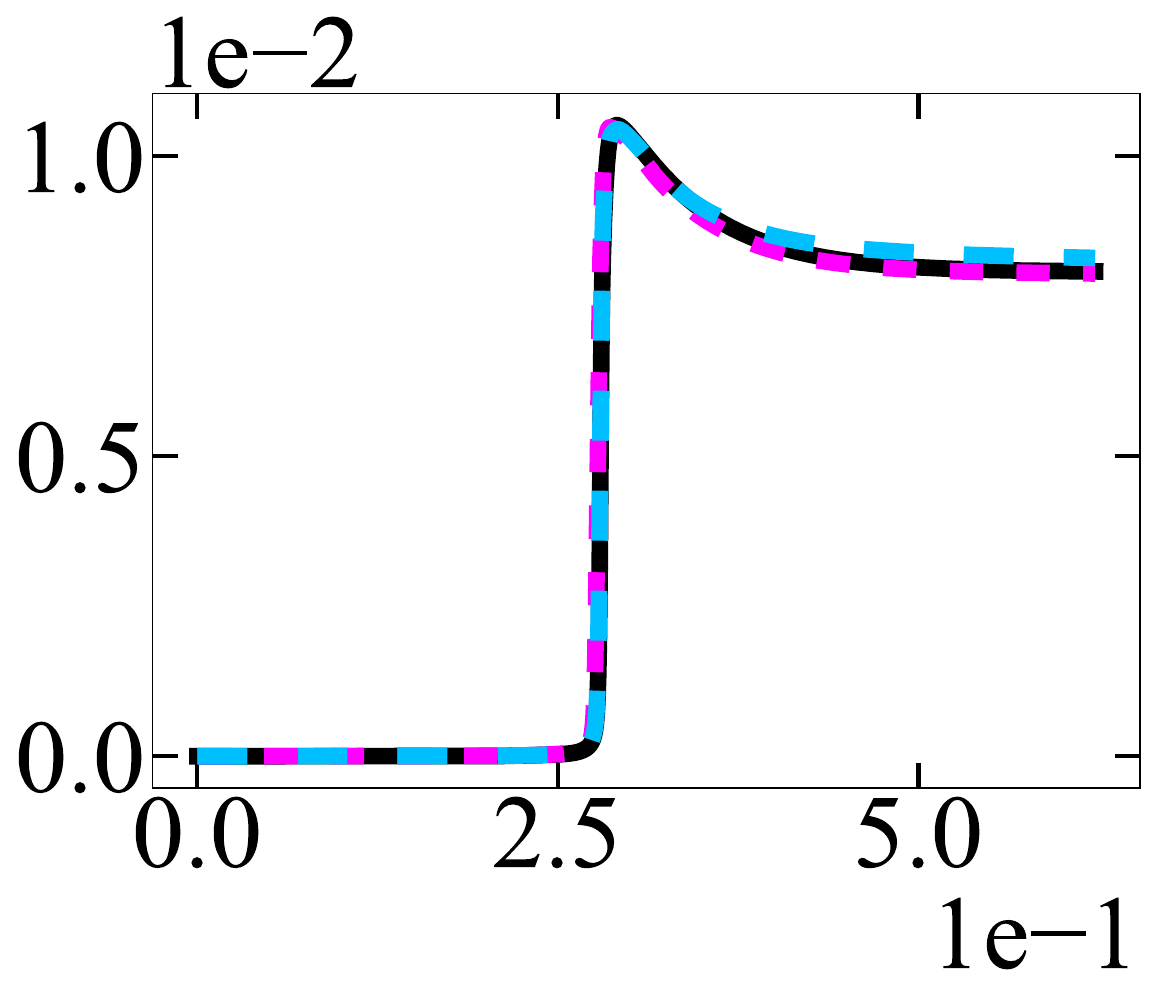}
\begin{picture}(0,0)
        \centering
       \put(\xlabelx, \xlabely){\scriptsize t (ms)}
        \put(\ylabelx, \ylabely){\makebox(0,0){\footnotesize{\rotatebox{90}{$Y_{OH}$}}}}
    \end{picture}
\hspace{\figspacex}
\includegraphics[width=\figwidth, height=\figheight]{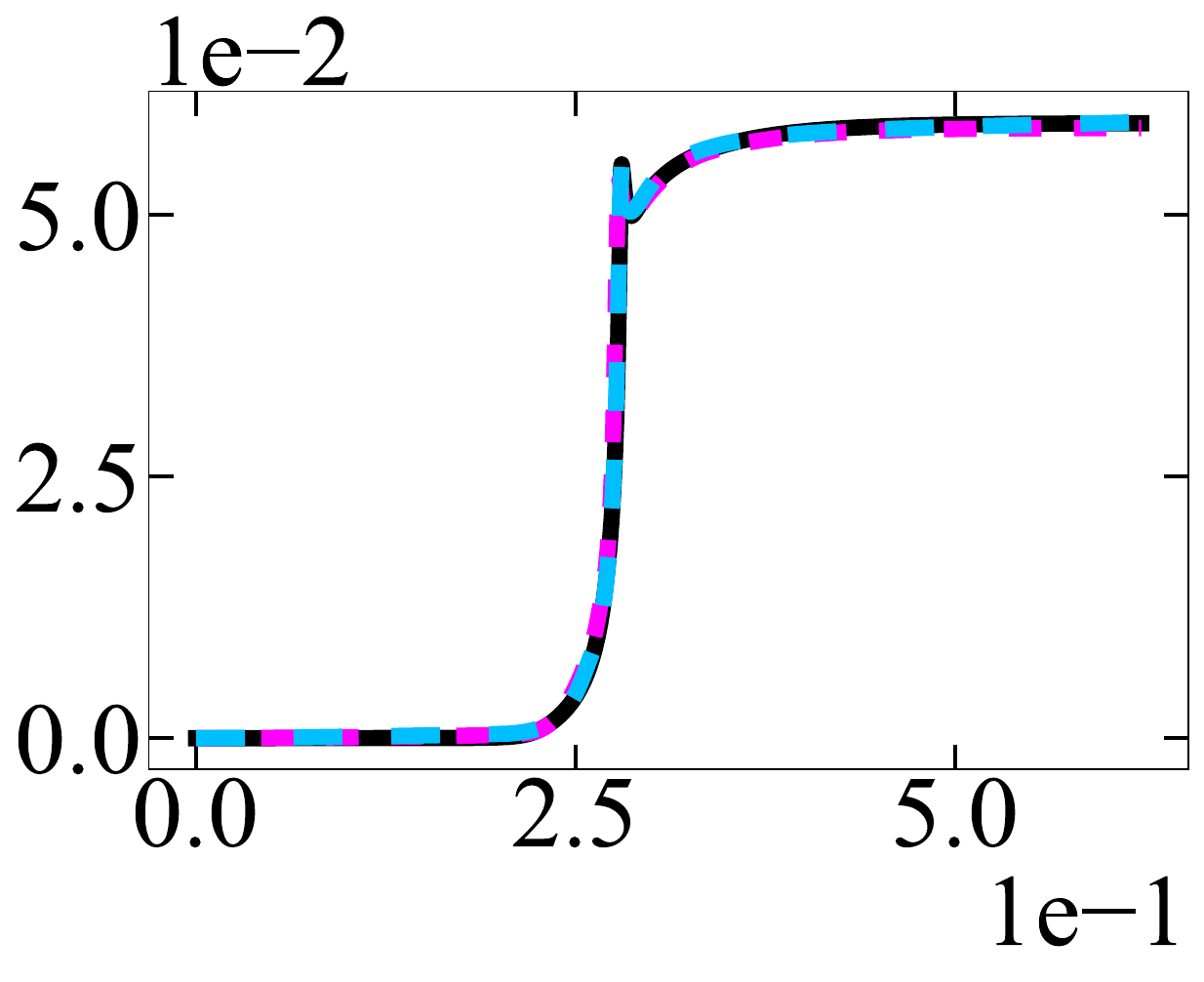}
\begin{picture}(0,0)
        \centering
       \put(\xlabelx, \xlabely){\scriptsize t (ms)}
        \put(\ylabelx, \ylabely){\makebox(0,0){\footnotesize{\rotatebox{90}{$Y_{H_2O}$}}}}
    \end{picture}
\hspace{\figspacex}
\includegraphics[width=\figwidth, height=\figheight]{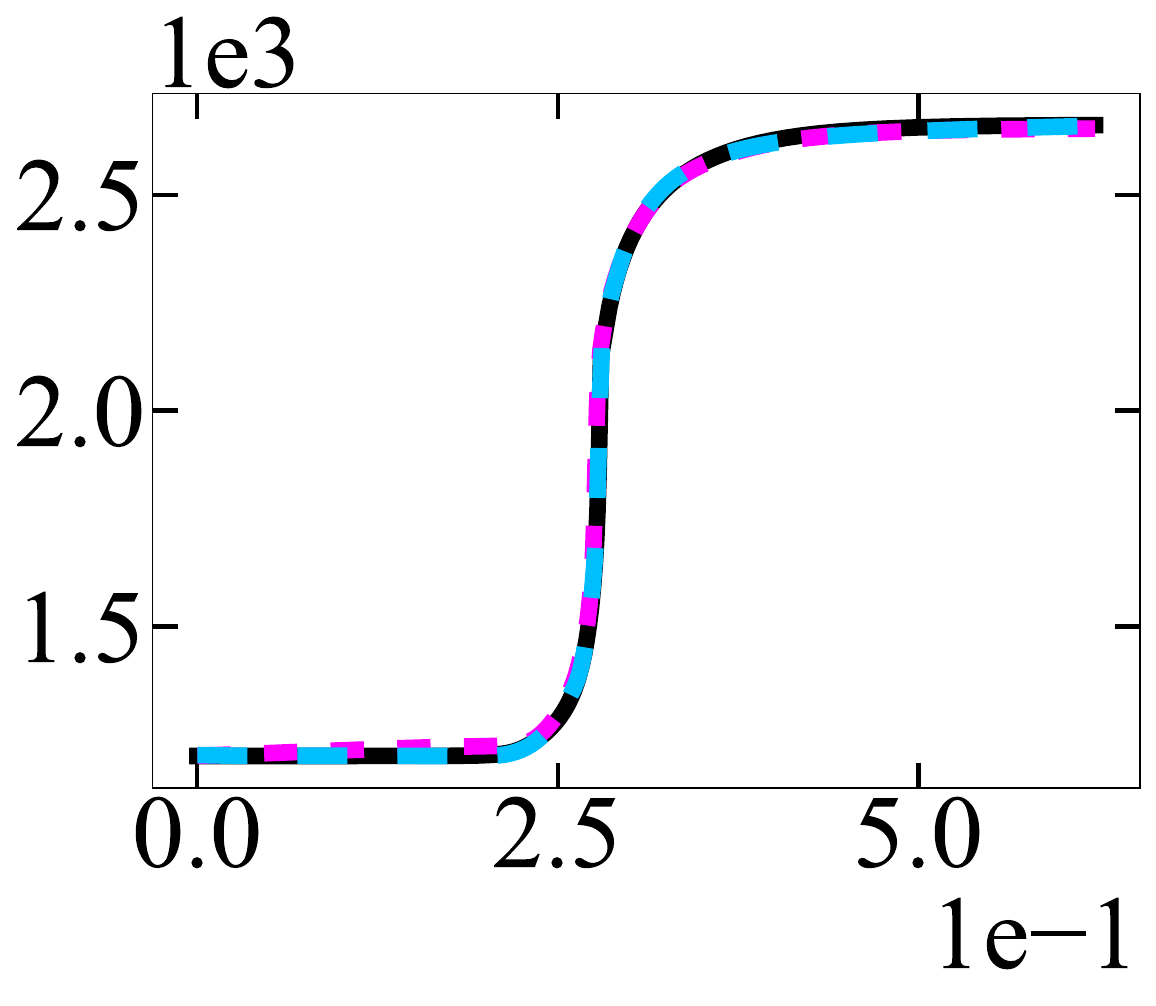}
\begin{picture}(0,0)
        \centering
       \put(\xlabelx, \xlabely){\scriptsize t (ms)}
        \put(\ylabelx, \ylabely){\makebox(0,0){\footnotesize{\rotatebox{90}{$T$}}}}
    \end{picture}
\hspace{\figspacex}
\includegraphics[width=\figwidth, height=\figheight]{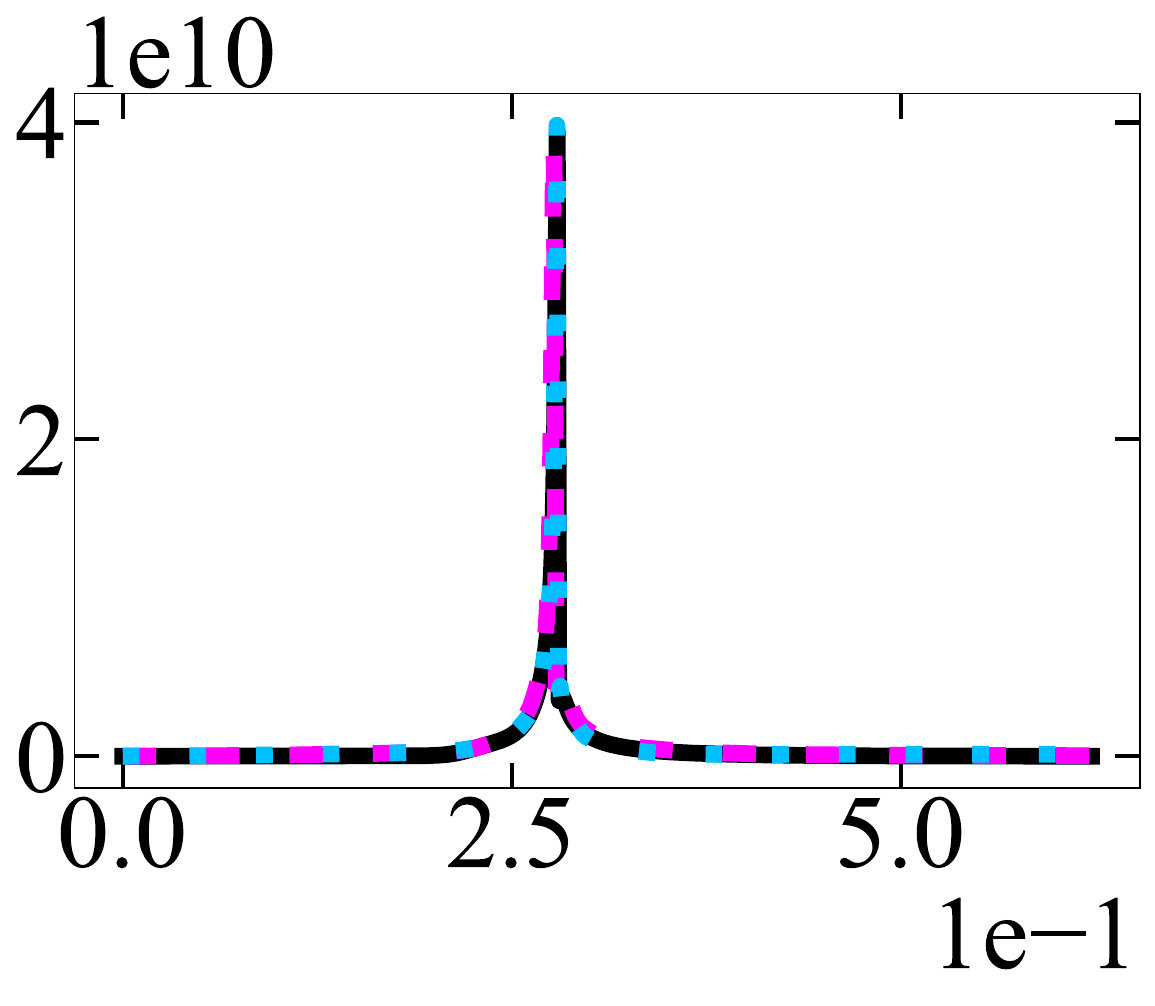}
\begin{picture}(0,0)
        \centering
       \put(\xlabelx, \xlabely){\scriptsize t (ms)}
        \put(\ylabelx, \ylabely){\makebox(0,0){\footnotesize{\rotatebox{90}{$HRR$}}}}
    \end{picture}

\vspace{0.1cm}
\caption{\footnotesize Temporal profiles of mass fractions of $C_2H_4$, $OH$, and $H_2O$, temperature ($T$) and heat release rate ($HRR$). (Black) actual profiles computed using Cantera. (Magenta) and (Cyan) are reconstructed profiles based on {\pca} and {\cokpca} LDMs, respectively.}
\label{YTs node ethylene}
\end{figure*}

 As previously mentioned, the {\node} solver can mitigate the propagation of errors over time. Accordingly, the PC profiles computed using the {\node} solver are nondivergent relative to the \textit{a priori} profiles. Figure~\ref{fig:PCs_node_ethylene} illustrates the evolution of PC profiles for a test configuration. Good agreement is observed between the \textit{a priori} and {\node} profiles for both the {\pca} and {\cokpca} based LDMs. The first two PCs are identical for both {\pca} and {\cokpca}. The subsequent PCs are different, with {\pca} profiles exhibiting a greater degree of stiffness. Minor departures from \textit{a priori} profiles are observed earlier in both cases for some of the PCs. To further assess the quality of the LDMs, we reconstructed thermochemical scalars from the PCs using separate ANNs. Figure~\ref{YTs node ethylene} shows some of the reconstructed species, temperature, and heat release rate computed from the thermochemical scalars. Again, the results from both {\pca} and {\cokpca} are in very good agreement with the actual profiles for all quantities shown. The performance of the {\cokpca} based LDM relative to {\pca} is visually indistinguishable from these plots. Therefore, we compute the cumulative errors in time to assess the relative performance. We quantify the error for a quantity $u$ as 
\begin{equation*} 
    \epsilon(u) = \frac{1}{n}\frac{\sum_{i=1}^{n} |u_{pred}(t_i)-u(t_i)|}{\max\limits_{i} \; |u(t_i)|}. 
\end{equation*} 
Table~\ref{errors entire ethylene} presents the cumulative errors over the entire time interval for the ten test configurations. It appears that {\pca} performs better than {\cokpca}, particularly in terms of capturing mass fractions. This is expected because the {\pca}-based LDM optimizes the errors for the majority samples. To assess the performance within the ignition zone, Tab.~\ref{errors ignition ethylene} shows the cumulative errors computed in the progress variable (based on the temperature) range $[0.05,0.95]$. While the reactant ($C_2H_4$) and product ($H_2O$) species are captured better with {\pca}, all minor species are better represented by {\cokpca}. Furthermore, {\cokpca} outperformed in minimizing errors in the heat release rate, which is a nonlinear aggregated quantity. These observations that {\cokpca} performs better than {\pca} in the reaction zone are consistent with the results of \textit{a priori} analyses presented in \cite{jonnalagadda2023co,nayak2024co}.

\begin{table*}[h] \scriptsize
\centering

\newcolumntype{?}{!{\vrule width 0pt}}
\begin{NiceTabular}{?c?c?c?c?c?c?c?c?c?c?c?c?}
\hline
\multicolumn{2}{?c?}{\cellcolor{gray!20}\textbf{T}} & \Block[fill=gray!20]{}{\textbf{1271}} & \Block[fill=gray!20]{}{\textbf{1146}} & \Block[fill=gray!20]{}{\textbf{1207}} & \Block[fill=gray!20]{}{\textbf{1242}} & \Block[fill=gray!20]{}{\textbf{1100}} & \Block[fill=gray!20]{}{\textbf{1175}} & \Block[fill=gray!20]{}{\textbf{1224}} & \Block[fill=gray!20]{}{\textbf{1300}} & \Block[fill=gray!20]{}{\textbf{1200}} & \Block[fill=gray!20]{}{\textbf{1247}} \\ \hline
\multicolumn{2}{?c?}{\cellcolor{gray!20}$\boldsymbol{\phi}$} & \Block[fill=gray!20]{}{\textbf{0.54}} & \Block[fill=gray!20]{}{\textbf{0.5}} & \Block[fill=gray!20]{}{\textbf{0.61}} & \Block[fill=gray!20]{}{\textbf{0.69}} & \Block[fill=gray!20]{}{\textbf{0.6}} & \Block[fill=gray!20]{}{\textbf{0.75}} & \Block[fill=gray!20]{}{\textbf{0.75}} & \Block[fill=gray!20]{}{\textbf{0.79}} & \Block[fill=gray!20]{}{\textbf{0.8}} & \Block[fill=gray!20]{}{\textbf{0.9}} \\ \hline
\Block[fill=gray!20]{2-1}{\textbf{$\mathbf{H_2O}$}} & \Block[fill=Lavender]{}{\textbf{PCA}} & \Block[fill=green!35]{}{3e-03} & \Block[fill=green!35]{}{2e-02} & \Block[fill=green!35]{}{3e-03} & \Block[fill=green!35]{}{5e-03} & \Block[fill=green!35]{}{2e-02} & \Block[fill=green!35]{}{4e-03} & \Block[fill=green!35]{}{7e-03} & \Block[fill=green!35]{}{3e-03} & \Block[fill=white]{}{7e-03} & \Block[fill=white]{}{6e-03}\\ 
 & \Block[fill=Apricot!35]{}{\textbf{COK}} & \Block[fill=white]{}{8e-03} & \Block[fill=white]{}{9e-02} & \Block[fill=white]{}{7e-03} & \Block[fill=white]{}{1e-02} & \Block[fill=white]{}{2e-02} & \Block[fill=white]{}{1e-02} & \Block[fill=white]{}{1e-02} & \Block[fill=white]{}{3e-03} & \Block[fill=green!35]{}{4e-03} & \Block[fill=green!35]{}{4e-03} \\ \hline
\Block[fill=gray!20]{2-1}{\textbf{$\mathbf{C_2H_4}$}} & \Block[fill=Lavender]{}{\textbf{PCA}} & \Block[fill=green!35]{}{2e-03} & \Block[fill=white]{}{1e-02} & \Block[fill=white]{}{4e-03} & \Block[fill=white]{}{5e-03} & \Block[fill=green!35]{}{4e-03} & \Block[fill=green!35]{}{3e-03} & \Block[fill=white]{}{3e-03} & \Block[fill=white]{}{2e-03} & \Block[fill=white]{}{6e-03} & \Block[fill=green!35]{}{4e-03}\\ 
 & \Block[fill=Apricot!35]{}{\textbf{COK}} & \Block[fill=white]{}{3e-03} & \Block[fill=green!35]{}{6e-03} & \Block[fill=green!35]{}{2e-03} & \Block[fill=green!35]{}{2e-03} & \Block[fill=white]{}{2e-02} & \Block[fill=white]{}{4e-03} & \Block[fill=green!35]{}{3e-03} & \Block[fill=green!35]{}{1e-03} & \Block[fill=green!35]{}{1e-03} & \Block[fill=white]{}{5e-03} \\ \hline
\Block[fill=gray!20]{2-1}{\textbf{$\mathbf{OH}$}} & \Block[fill=Lavender]{}{\textbf{PCA}} & \Block[fill=green!35]{}{3e-03} & \Block[fill=green!35]{}{2e-02} & \Block[fill=green!35]{}{2e-03} & \Block[fill=white]{}{7e-03} & \Block[fill=green!35]{}{1e-02} & \Block[fill=green!35]{}{3e-03} & \Block[fill=green!35]{}{5e-03} & \Block[fill=green!35]{}{4e-03} & \Block[fill=green!35]{}{6e-03} & \Block[fill=white]{}{1e-02}\\ 
 & \Block[fill=Apricot!35]{}{\textbf{COK}} & \Block[fill=white]{}{7e-03} & \Block[fill=white]{}{5e-02} & \Block[fill=white]{}{9e-03} & \Block[fill=green!35]{}{7e-03} & \Block[fill=white]{}{3e-02} & \Block[fill=white]{}{6e-03} & \Block[fill=white]{}{2e-02} & \Block[fill=white]{}{6e-03} & \Block[fill=white]{}{1e-02} & \Block[fill=green!35]{}{5e-03} \\ \hline
\Block[fill=gray!20]{2-1}{\textbf{T}} & \Block[fill=Lavender]{}{\textbf{PCA}} & \Block[fill=green!35]{}{2e-03} & \Block[fill=white]{}{9e-03} & \Block[fill=white]{}{2e-03} & \Block[fill=green!35]{}{3e-03} & \Block[fill=white]{}{2e-02} & \Block[fill=white]{}{3e-03} & \Block[fill=green!35]{}{4e-03} & \Block[fill=green!35]{}{2e-03} & \Block[fill=white]{}{4e-03} & \Block[fill=white]{}{4e-03}\\ 
 & \Block[fill=Apricot!35]{}{\textbf{COK}} & \Block[fill=white]{}{3e-03} & \Block[fill=green!35]{}{5e-03} & \Block[fill=green!35]{}{1e-03} & \Block[fill=white]{}{4e-03} & \Block[fill=green!35]{}{8e-03} & \Block[fill=green!35]{}{3e-03} & \Block[fill=white]{}{5e-03} & \Block[fill=white]{}{2e-03} & \Block[fill=green!35]{}{1e-03} & \Block[fill=green!35]{}{2e-03} \\ \hline
\Block[fill=gray!20]{2-1}{\textbf{HRR}} & \Block[fill=Lavender]{}{\textbf{PCA}} & \Block[fill=green!35]{}{5e-03} & \Block[fill=white]{}{9e-03} & \Block[fill=green!35]{}{4e-03} & \Block[fill=white]{}{8e-03} & \Block[fill=green!35]{}{9e-03} & \Block[fill=green!35]{}{2e-03} & \Block[fill=green!35]{}{5e-03} & \Block[fill=white]{}{7e-03} & \Block[fill=white]{}{6e-03} & \Block[fill=white]{}{9e-03}\\ 
 & \Block[fill=Apricot!35]{}{\textbf{COK}} & \Block[fill=white]{}{5e-03} & \Block[fill=green!35]{}{9e-03} & \Block[fill=white]{}{5e-03} & \Block[fill=green!35]{}{3e-03} & \Block[fill=white]{}{1e-02} & \Block[fill=white]{}{5e-03} & \Block[fill=white]{}{7e-03} & \Block[fill=green!35]{}{5e-03} & \Block[fill=green!35]{}{3e-03} & \Block[fill=green!35]{}{6e-03} \\ \hline
\end{NiceTabular}

\vspace{0.1cm}
\caption{Cumulative errors ($\epsilon$) in the entire time interval for test data with {\node} solver for ethylene-air mixture.}
\label{errors entire ethylene}

\end{table*}

\begin{table*}[h!] \scriptsize
\centering

\newcolumntype{?}{!{\vrule width 0pt}}
\begin{NiceTabular}{?c?c?c?c?c?c?c?c?c?c?c?c?}
\hline
\multicolumn{2}{?c?}{\cellcolor{gray!20}\textbf{T}} & \Block[fill=gray!20]{}{\textbf{1271}} & \Block[fill=gray!20]{}{\textbf{1146}} & \Block[fill=gray!20]{}{\textbf{1207}} & \Block[fill=gray!20]{}{\textbf{1242}} & \Block[fill=gray!20]{}{\textbf{1100}} & \Block[fill=gray!20]{}{\textbf{1175}} & \Block[fill=gray!20]{}{\textbf{1224}} & \Block[fill=gray!20]{}{\textbf{1300}} & \Block[fill=gray!20]{}{\textbf{1200}} & \Block[fill=gray!20]{}{\textbf{1247}} \\ \hline
\multicolumn{2}{?c?}{\cellcolor{gray!20}$\boldsymbol{\phi}$} & \Block[fill=gray!20]{}{\textbf{0.54}} & \Block[fill=gray!20]{}{\textbf{0.5}} & \Block[fill=gray!20]{}{\textbf{0.61}} & \Block[fill=gray!20]{}{\textbf{0.69}} & \Block[fill=gray!20]{}{\textbf{0.6}} & \Block[fill=gray!20]{}{\textbf{0.75}} & \Block[fill=gray!20]{}{\textbf{0.75}} & \Block[fill=gray!20]{}{\textbf{0.79}} & \Block[fill=gray!20]{}{\textbf{0.8}} & \Block[fill=gray!20]{}{\textbf{0.9}} \\ \hline
\Block[fill=gray!20]{2-1}{\textbf{$\mathbf{H_2O}$}} & \Block[fill=Lavender]{}{\textbf{PCA}} & \Block[fill=green!35]{}{4e-03} & \Block[fill=green!35]{}{2e-02} & \Block[fill=green!35]{}{6e-03} & \Block[fill=green!35]{}{6e-03} & \Block[fill=white]{}{2e-02} & \Block[fill=green!35]{}{7e-03} & \Block[fill=white]{}{9e-03} & \Block[fill=green!35]{}{5e-03} & \Block[fill=white]{}{9e-03} & \Block[fill=white]{}{8e-03}\\ 
 & \Block[fill=Apricot!35]{}{\textbf{COK}} & \Block[fill=white]{}{9e-03} & \Block[fill=white]{}{2e-02} & \Block[fill=white]{}{9e-03} & \Block[fill=white]{}{1e-02} & \Block[fill=green!35]{}{5e-03} & \Block[fill=white]{}{8e-03} & \Block[fill=green!35]{}{5e-03} & \Block[fill=white]{}{5e-03} & \Block[fill=green!35]{}{6e-03} & \Block[fill=green!35]{}{4e-03} \\ \hline
\Block[fill=gray!20]{2-1}{\textbf{$\mathbf{C_2H_4}$}} & \Block[fill=Lavender]{}{\textbf{PCA}} & \Block[fill=green!35]{}{3e-03} & \Block[fill=white]{}{2e-02} & \Block[fill=green!35]{}{4e-03} & \Block[fill=white]{}{3e-03} & \Block[fill=white]{}{3e-02} & \Block[fill=green!35]{}{5e-03} & \Block[fill=white]{}{3e-03} & \Block[fill=green!35]{}{3e-03} & \Block[fill=white]{}{4e-03} & \Block[fill=white]{}{3e-03}\\ 
 & \Block[fill=Apricot!35]{}{\textbf{COK}} & \Block[fill=white]{}{3e-03} & \Block[fill=green!35]{}{1e-02} & \Block[fill=white]{}{4e-03} & \Block[fill=green!35]{}{2e-03} & \Block[fill=green!35]{}{4e-03} & \Block[fill=white]{}{6e-03} & \Block[fill=green!35]{}{3e-03} & \Block[fill=white]{}{6e-03} & \Block[fill=green!35]{}{2e-03} & \Block[fill=green!35]{}{3e-03} \\ \hline
\Block[fill=gray!20]{2-1}{\textbf{$\mathbf{OH}$}} & \Block[fill=Lavender]{}{\textbf{PCA}} & \Block[fill=white]{}{7e-03} & \Block[fill=green!35]{}{8e-03} & \Block[fill=green!35]{}{5e-03} & \Block[fill=green!35]{}{5e-03} & \Block[fill=white]{}{9e-03} & \Block[fill=green!35]{}{5e-03} & \Block[fill=white]{}{9e-03} & \Block[fill=green!35]{}{8e-03} & \Block[fill=white]{}{9e-03} & \Block[fill=white]{}{3e-02}\\ 
 & \Block[fill=Apricot!35]{}{\textbf{COK}} & \Block[fill=green!35]{}{6e-03} & \Block[fill=white]{}{2e-02} & \Block[fill=white]{}{8e-03} & \Block[fill=white]{}{1e-02} & \Block[fill=green!35]{}{7e-03} & \Block[fill=white]{}{5e-03} & \Block[fill=green!35]{}{4e-03} & \Block[fill=white]{}{1e-02} & \Block[fill=green!35]{}{3e-03} & \Block[fill=green!35]{}{4e-03} \\ \hline
\Block[fill=gray!20]{2-1}{\textbf{T}} & \Block[fill=Lavender]{}{\textbf{PCA}} & \Block[fill=green!35]{}{3e-03} & \Block[fill=green!35]{}{5e-03} & \Block[fill=white]{}{3e-03} & \Block[fill=green!35]{}{2e-03} & \Block[fill=white]{}{9e-03} & \Block[fill=white]{}{3e-03} & \Block[fill=white]{}{4e-03} & \Block[fill=green!35]{}{2e-03} & \Block[fill=white]{}{4e-03} & \Block[fill=white]{}{6e-03}\\ 
 & \Block[fill=Apricot!35]{}{\textbf{COK}} & \Block[fill=white]{}{4e-03} & \Block[fill=white]{}{8e-03} & \Block[fill=green!35]{}{2e-03} & \Block[fill=white]{}{4e-03} & \Block[fill=green!35]{}{2e-03} & \Block[fill=green!35]{}{2e-03} & \Block[fill=green!35]{}{2e-03} & \Block[fill=white]{}{5e-03} & \Block[fill=green!35]{}{2e-03} & \Block[fill=green!35]{}{1e-03} \\ \hline
\Block[draw=magenta,line-width=1.5pt]{10-12}{} \Block[fill=gray!20]{2-1}{\textbf{$\mathbf{HO_2}$}} & \Block[fill=Lavender]{}{\textbf{PCA}} & \Block[fill=white]{}{4e-03} & \Block[fill=green!35]{}{7e-03} & \Block[fill=green!35]{}{4e-03} & \Block[fill=white]{}{4e-03} & \Block[fill=white]{}{2e-02} & \Block[fill=white]{}{4e-03} & \Block[fill=white]{}{7e-03} & \Block[fill=white]{}{8e-03} & \Block[fill=white]{}{9e-03} & \Block[fill=white]{}{9e-03}\\ 
 & \Block[fill=Apricot!35]{}{\textbf{COK}} & \Block[fill=green!35]{}{4e-03} & \Block[fill=white]{}{1e-02} & \Block[fill=white]{}{5e-03} & \Block[fill=green!35]{}{3e-03} & \Block[fill=green!35]{}{1e-02} & \Block[fill=green!35]{}{3e-03} & \Block[fill=green!35]{}{3e-03} & \Block[fill=green!35]{}{4e-03} & \Block[fill=green!35]{}{4e-03} & \Block[fill=green!35]{}{3e-03} \\ \hline
\Block[fill=gray!20]{2-1}{\textbf{$\mathbf{CH_2O}$}} & \Block[fill=Lavender]{}{\textbf{PCA}} & \Block[fill=green!35]{}{3e-03} & \Block[fill=white]{}{8e-03} & \Block[fill=white]{}{3e-03} & \Block[fill=white]{}{3e-03} & \Block[fill=white]{}{1e-02} & \Block[fill=white]{}{5e-03} & \Block[fill=green!35]{}{3e-03} & \Block[fill=white]{}{7e-03} & \Block[fill=white]{}{5e-03} & \Block[fill=white]{}{4e-03}\\ 
 & \Block[fill=Apricot!35]{}{\textbf{COK}} & \Block[fill=white]{}{3e-03} & \Block[fill=green!35]{}{6e-03} & \Block[fill=green!35]{}{3e-03} & \Block[fill=green!35]{}{2e-03} & \Block[fill=green!35]{}{3e-03} & \Block[fill=green!35]{}{4e-03} & \Block[fill=white]{}{4e-03} & \Block[fill=green!35]{}{4e-03} & \Block[fill=green!35]{}{3e-03} & \Block[fill=green!35]{}{3e-03} \\ \hline
\Block[fill=gray!20]{2-1}{\textbf{$\mathbf{CH_4}$}} & \Block[fill=Lavender]{}{\textbf{PCA}} & \Block[fill=white]{}{5e-03} & \Block[fill=white]{}{1e-02} & \Block[fill=white]{}{5e-03} & \Block[fill=white]{}{4e-03} & \Block[fill=white]{}{3e-02} & \Block[fill=white]{}{6e-03} & \Block[fill=white]{}{4e-03} & \Block[fill=white]{}{8e-03} & \Block[fill=white]{}{6e-03} & \Block[fill=white]{}{5e-03}\\ 
 & \Block[fill=Apricot!35]{}{\textbf{COK}} & \Block[fill=green!35]{}{3e-03} & \Block[fill=green!35]{}{1e-02} & \Block[fill=green!35]{}{5e-03} & \Block[fill=green!35]{}{2e-03} & \Block[fill=green!35]{}{3e-03} & \Block[fill=green!35]{}{5e-03} & \Block[fill=green!35]{}{3e-03} & \Block[fill=green!35]{}{5e-03} & \Block[fill=green!35]{}{4e-03} & \Block[fill=green!35]{}{3e-03} \\ \hline
\Block[fill=gray!20]{2-1}{\textbf{$\mathbf{CH}$}} & \Block[fill=Lavender]{}{\textbf{PCA}} & \Block[fill=white]{}{3e-03} & \Block[fill=green!35]{}{3e-03} & \Block[fill=white]{}{3e-03} & \Block[fill=white]{}{3e-03} & \Block[fill=white]{}{4e-03} & \Block[fill=white]{}{4e-03} & \Block[fill=white]{}{4e-03} & \Block[fill=green!35]{}{6e-03} & \Block[fill=white]{}{4e-03} & \Block[fill=white]{}{5e-03}\\ 
 & \Block[fill=Apricot!35]{}{\textbf{COK}} & \Block[fill=green!35]{}{2e-03} & \Block[fill=white]{}{6e-03} & \Block[fill=green!35]{}{3e-03} & \Block[fill=green!35]{}{2e-03} & \Block[fill=green!35]{}{2e-03} & \Block[fill=green!35]{}{2e-03} & \Block[fill=green!35]{}{2e-03} & \Block[fill=white]{}{6e-03} & \Block[fill=green!35]{}{3e-03} & \Block[fill=green!35]{}{5e-03} \\ \hline
\Block[fill=gray!20]{2-1}{\textbf{$\mathbf{H_2O_2}$}} & \Block[fill=Lavender]{}{\textbf{PCA}} & \Block[fill=white]{}{4e-03} & \Block[fill=white]{}{1e-02} & \Block[fill=white]{}{5e-03} & \Block[fill=white]{}{4e-03} & \Block[fill=white]{}{3e-02} & \Block[fill=green!35]{}{5e-03} & \Block[fill=white]{}{4e-03} & \Block[fill=white]{}{7e-03} & \Block[fill=white]{}{5e-03} & \Block[fill=white]{}{7e-03}\\ 
 & \Block[fill=Apricot!35]{}{\textbf{COK}} & \Block[fill=green!35]{}{4e-03} & \Block[fill=green!35]{}{1e-02} & \Block[fill=green!35]{}{4e-03} & \Block[fill=green!35]{}{2e-03} & \Block[fill=green!35]{}{1e-02} & \Block[fill=white]{}{5e-03} & \Block[fill=green!35]{}{3e-03} & \Block[fill=green!35]{}{7e-03} & \Block[fill=green!35]{}{3e-03} & \Block[fill=green!35]{}{4e-03} \\ \hline
\Block[draw=red,line-width=1.5pt]{2-12}{} \Block[fill=gray!20]{2-1}{\textbf{HRR}} & \Block[fill=Lavender]{}{\textbf{PCA}} & \Block[fill=white]{}{7e-03} & \Block[fill=green!35]{}{7e-03} & \Block[fill=white]{}{8e-03} & \Block[fill=white]{}{7e-03} & \Block[fill=white]{}{1e-02} & \Block[fill=white]{}{8e-03} & \Block[fill=white]{}{1e-02} & \Block[fill=white]{}{1e-02} & \Block[fill=white]{}{1e-02} & \Block[fill=white]{}{2e-02}\\ 
 & \Block[fill=Apricot!35]{}{\textbf{COK}} & \Block[fill=green!35]{}{5e-03} & \Block[fill=white]{}{9e-03} & \Block[fill=green!35]{}{7e-03} & \Block[fill=green!35]{}{6e-03} & \Block[fill=green!35]{}{6e-03} & \Block[fill=green!35]{}{5e-03} & \Block[fill=green!35]{}{4e-03} & \Block[fill=green!35]{}{9e-03} & \Block[fill=green!35]{}{5e-03} & \Block[fill=green!35]{}{8e-03} \\ \hline
\end{NiceTabular}

\vspace{0.1cm}
\caption{Cumulative errors ($\epsilon$) in the ignition zone for test data with {\node} solver for ethylene-air mixture.}
\label{errors ignition ethylene}

\end{table*}

\subsection{Spontaneous ignition of \textit{n}heptane-air mixture
\label{subsec:nheptane_results}} \addvspace{10pt}

\begin{figure*}[h!]
\newcommand{\figwidth}{0.9in}
\newcommand{\figheight}{0.8in}
\newcommand{\xlabelx}{-40}
\newcommand{\xlabely}{-5}
\newcommand{\ylabelx}{-73}
\newcommand{\ylabely}{32}
\newcommand{\figspacex}{0.2cm}
\centering
\includegraphics[width=\figwidth, height=\figheight]{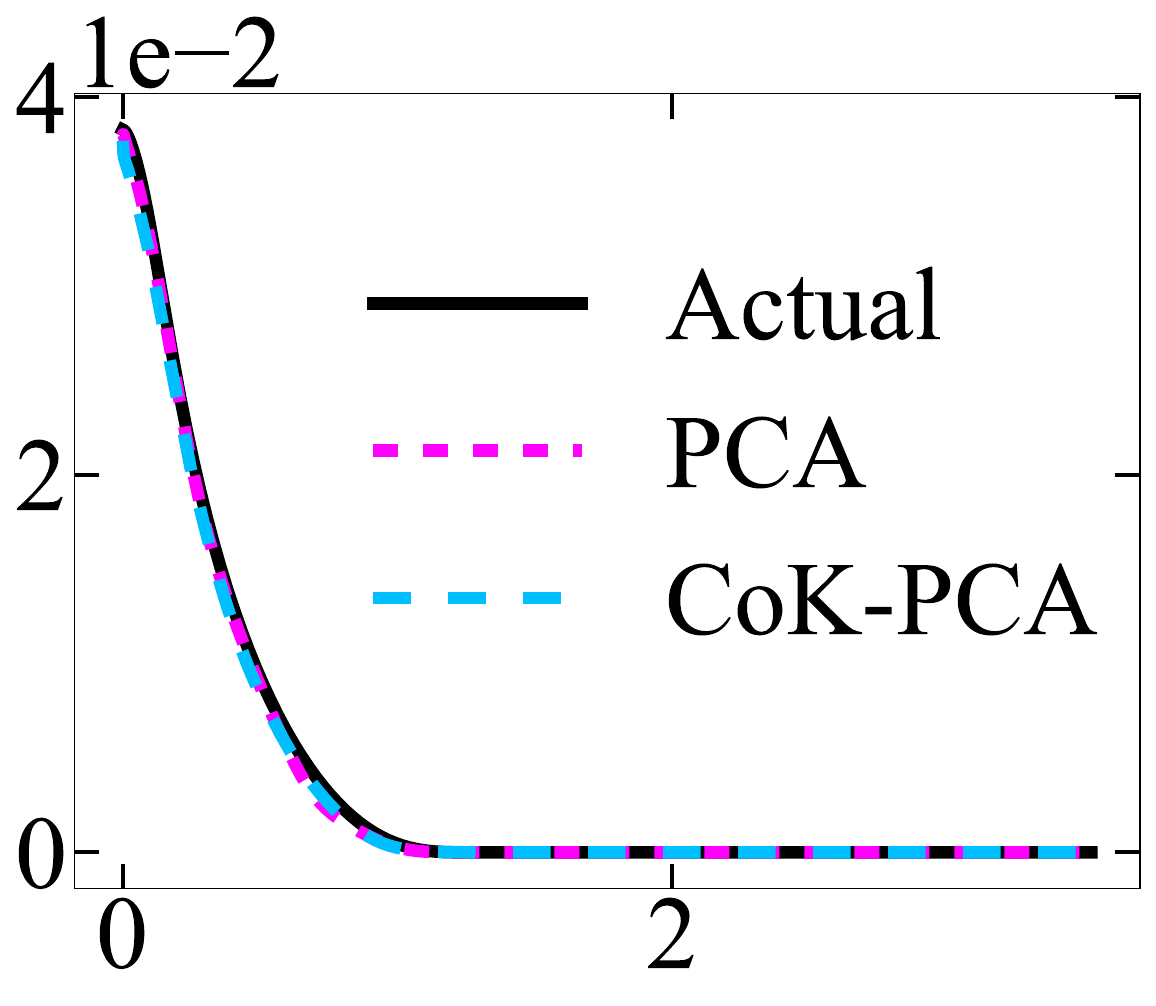}
\begin{picture}(0,0)
        \centering
       \put(\xlabelx, \xlabely){\scriptsize t (ms)}
        \put(\ylabelx, \ylabely){\makebox(0,0){\footnotesize{\rotatebox{90}{$Y_{C_7H_{16}}$}}}}
    \end{picture}
\hspace{\figspacex}
\begin{tikzpicture}
    \begin{scope}[spy using outlines={circle, magnification=1.6, size=0.8cm}]
        \node[inner sep=0pt] (main) at (0,0) {\includegraphics[width=\figwidth,height=\figheight]{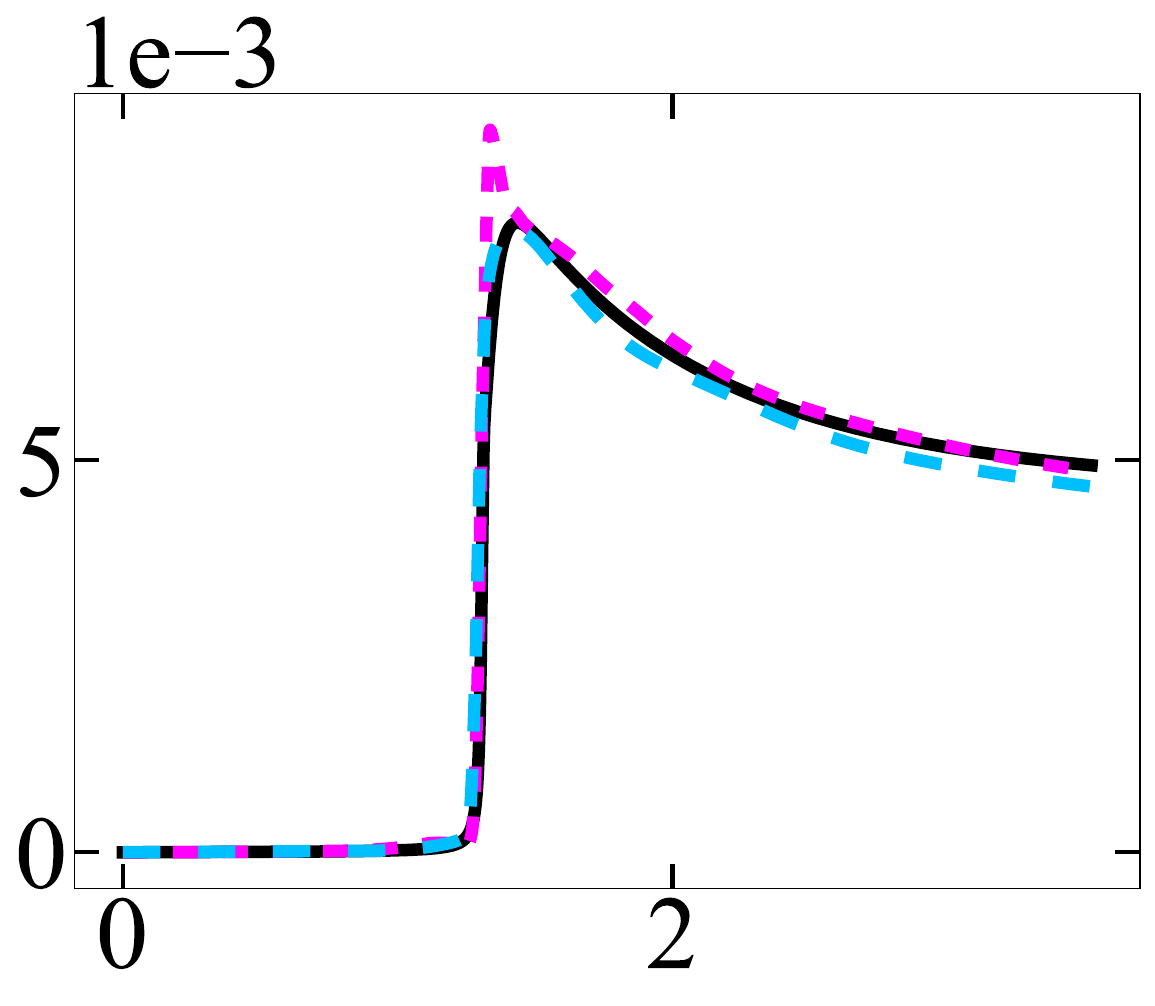}
        \begin{picture}(0,0)
            \centering
            \put(\xlabelx, \xlabely){\scriptsize t (ms)}
            \put(\ylabelx, \ylabely){\makebox(0,0){\footnotesize{\rotatebox{90}{$Y_{OH}$}}}}
        \end{picture}};
        \spy [red] on (-0.1,0.6) in node [right] at (0,-0.3);
    \end{scope}
    \draw[->,thin,red] (tikzspyonnode) -- (tikzspyinnode);
\end{tikzpicture}
\hspace{\figspacex}
\includegraphics[width=\figwidth, height=\figheight]{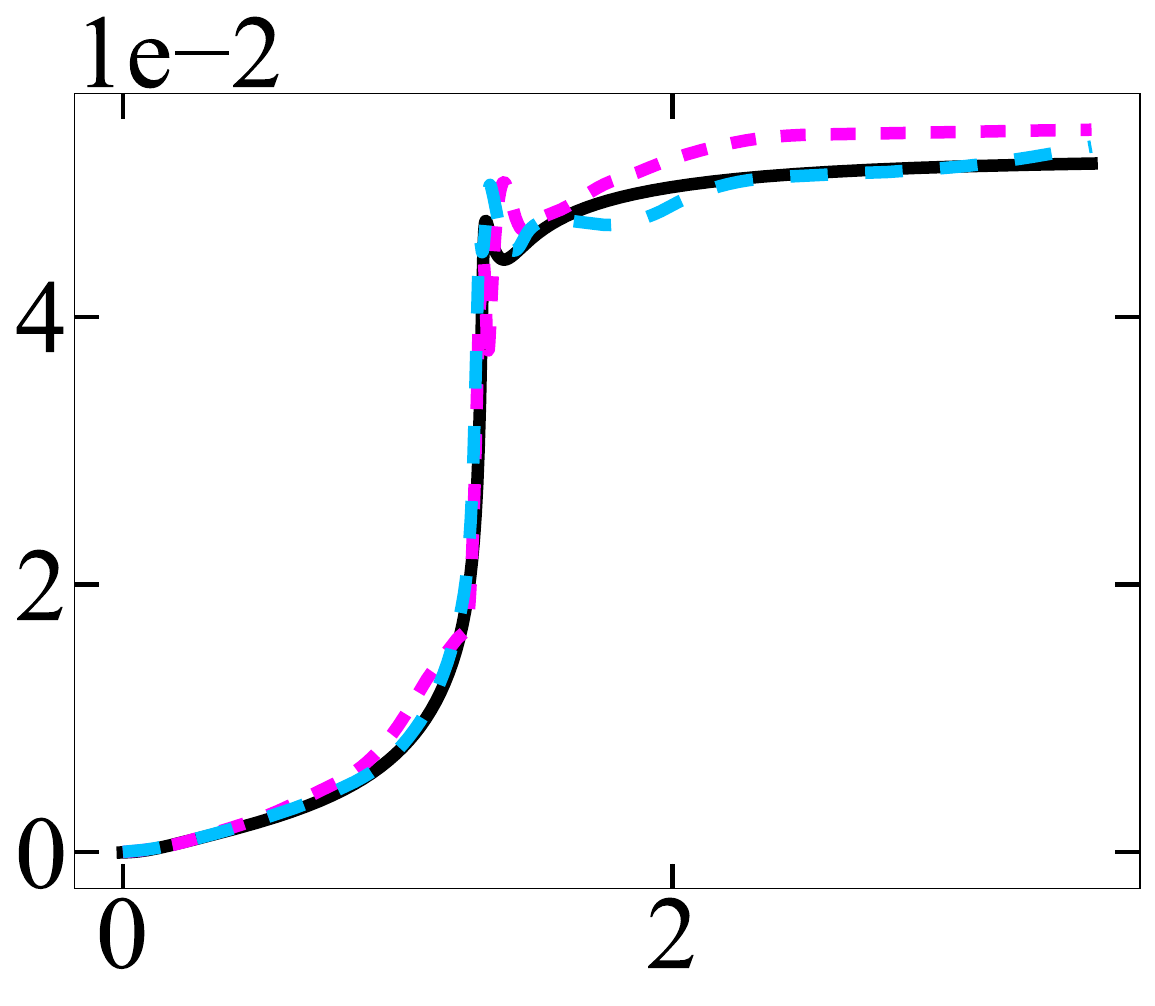}
\begin{picture}(0,0)
        \centering
       \put(\xlabelx, \xlabely){\scriptsize t (ms)}
        \put(\ylabelx, \ylabely){\makebox(0,0){\footnotesize{\rotatebox{90}{$Y_{H_2O}$}}}}
    \end{picture}
\hspace{\figspacex}
\begin{tikzpicture}
    \begin{scope}[spy using outlines={circle, magnification=1.6, size=0.8cm}]
        \node[inner sep=0pt] (main) at (0,0) {\includegraphics[width=\figwidth,height=\figheight]{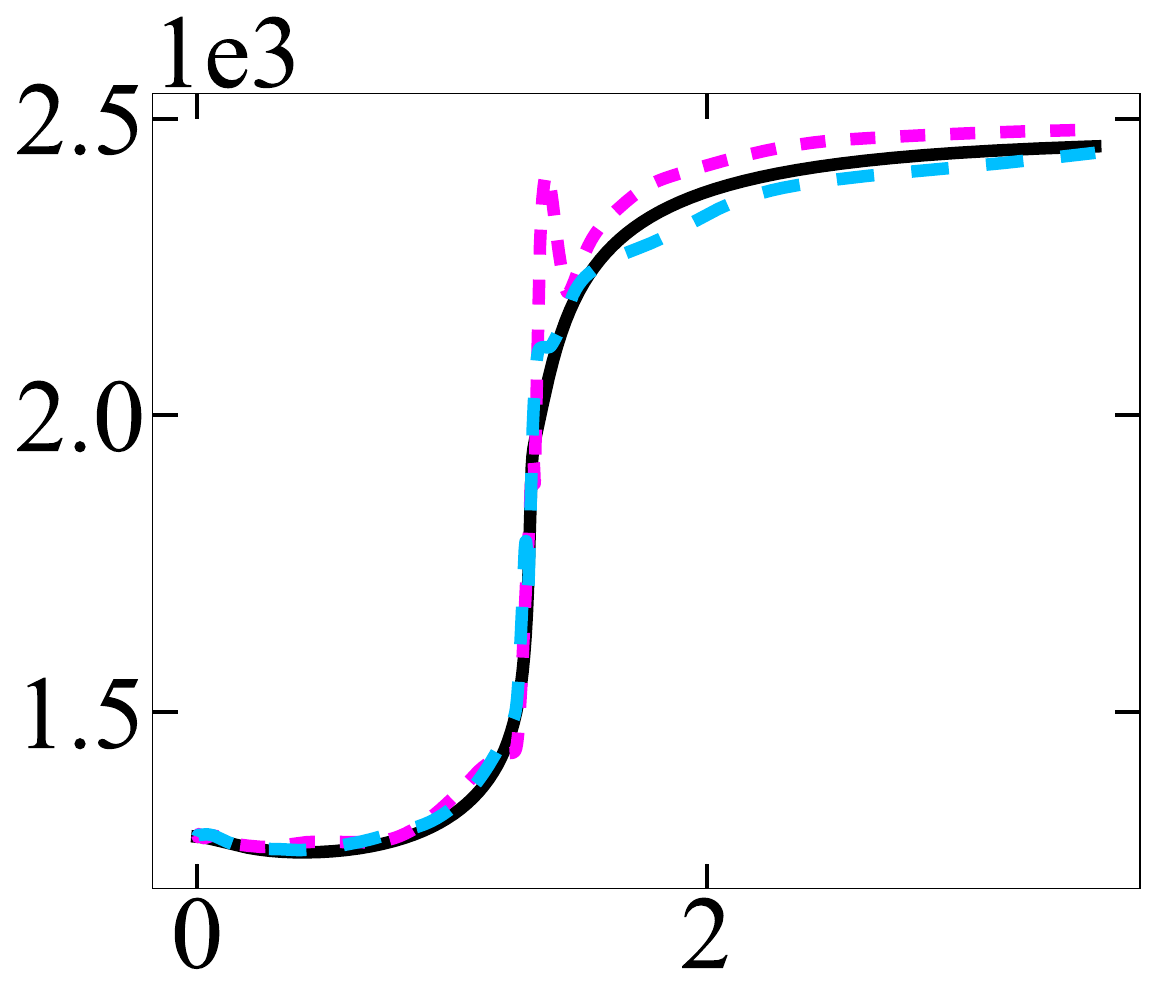}
        \begin{picture}(0,0)
            \centering
            \put(\xlabelx, \xlabely){\scriptsize t (ms)}
            \put(\ylabelx, \ylabely){\makebox(0,0){\footnotesize{\rotatebox{90}{$T$}}}}
        \end{picture}};
        \spy [red] on (-0.1,0.5) in node [right] at (0,-0.3);
    \end{scope}
    \draw[->,thin,red] (tikzspyonnode) -- (tikzspyinnode);
\end{tikzpicture}
\hspace{\figspacex}
\begin{tikzpicture}
    \begin{scope}[spy using outlines={circle, magnification=1.6, size=0.8cm}]
        \node[inner sep=0pt] (main) at (0,0) {\includegraphics[width=\figwidth,height=\figheight]{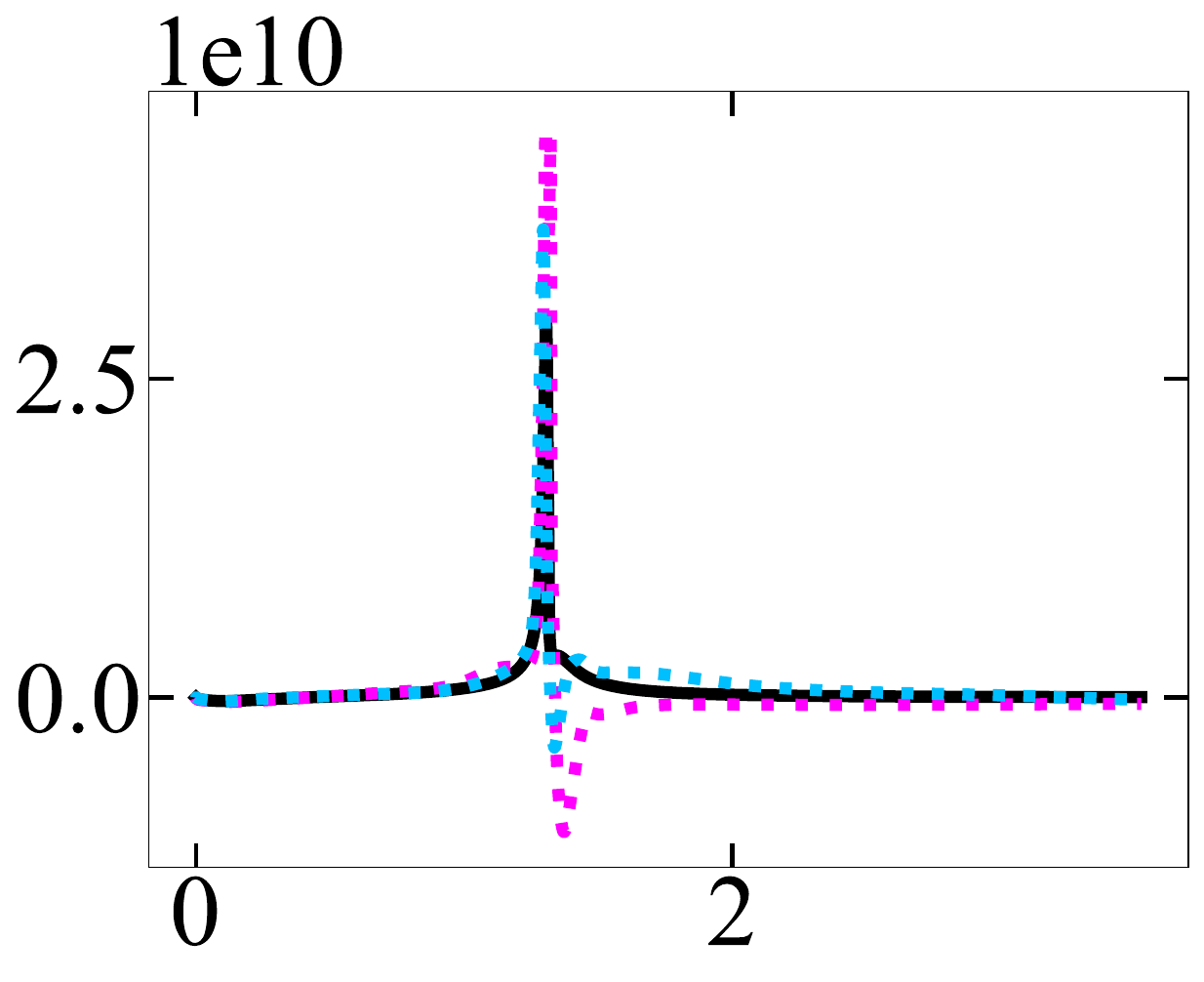}
        \begin{picture}(0,0)
            \centering
            \put(\xlabelx, \xlabely){\scriptsize t (ms)}
            \put(\ylabelx, \ylabely){\makebox(0,0){\footnotesize{\rotatebox{90}{$HRR$}}}}
        \end{picture}};
        \spy [red] on (-0.1,-0.5) in node [right] at (0,0.2);
    \end{scope}
    \draw[->,thin,red] (tikzspyonnode) -- (tikzspyinnode);
\end{tikzpicture}

\vspace{0.1cm}
\caption{\footnotesize Temporal profiles of mass fractions of \textit{n}heptane, $OH$, and $H_2O$, temperature ($T$) and heat release rate ($HRR$). (Black) actual profiles computed using Cantera. (Magenta) and (Cyan) are reconstructed profiles based on {\pca} and {\cokpca} LDMs, respectively.}
\label{YTs node heptane}
\end{figure*}

\begin{table*}[h!] \scriptsize
\centering
\newcolumntype{?}{!{\vrule width 0pt}}
\begin{NiceTabular}{?c?c?c?c?c?c?c?c?c?c?c?c?}
\hline
\multicolumn{2}{?c?}{\cellcolor{gray!20}\textbf{T}} & \Block[fill=gray!20]{}{\textbf{1129}} & \Block[fill=gray!20]{}{\textbf{1250}} & \Block[fill=gray!20]{}{\textbf{1207}} & \Block[fill=gray!20]{}{\textbf{1129}} & \Block[fill=gray!20]{}{\textbf{1102}} & \Block[fill=gray!20]{}{\textbf{1290}} & \Block[fill=gray!20]{}{\textbf{1190}} & \Block[fill=gray!20]{}{\textbf{1262}} & \Block[fill=gray!20]{}{\textbf{1150}} & \Block[fill=gray!20]{}{\textbf{1300}} \\ \hline
\multicolumn{2}{?c?}{\cellcolor{gray!20}$\boldsymbol{\phi}$} & \Block[fill=gray!20]{}{\textbf{0.56}} & \Block[fill=gray!20]{}{\textbf{0.58}} & \Block[fill=gray!20]{}{\textbf{0.59}} & \Block[fill=gray!20]{}{\textbf{0.64}} & \Block[fill=gray!20]{}{\textbf{0.6}} & \Block[fill=gray!20]{}{\textbf{0.6}} & \Block[fill=gray!20]{}{\textbf{0.78}} & \Block[fill=gray!20]{}{\textbf{0.82}} & \Block[fill=gray!20]{}{\textbf{0.8}} & \Block[fill=gray!20]{}{\textbf{0.8}} \\ \hline
\Block[fill=gray!20]{2-1}{\textbf{$\mathbf{H_2O}$}} & \Block[fill=Lavender]{}{\textbf{PCA}} & \Block[fill=white]{}{4e-02} & \Block[fill=white]{}{4e-02} & \Block[fill=green!35]{}{3e-02} & \Block[fill=white]{}{3e-02} & \Block[fill=green!35]{}{7e-02} & \Block[fill=white]{}{4e-02} & \Block[fill=white]{}{2e-02} & \Block[fill=white]{}{4e-02} & \Block[fill=white]{}{2e-02} & \Block[fill=white]{}{5e-02}\\ 
 & \Block[fill=Apricot!35]{}{\textbf{COK}} & \Block[fill=green!35]{}{2e-02} & \Block[fill=green!35]{}{2e-02} & \Block[fill=white]{}{4e-02} & \Block[fill=green!35]{}{1e-02} & \Block[fill=white]{}{2e-01} & \Block[fill=green!35]{}{2e-02} & \Block[fill=green!35]{}{1e-02} & \Block[fill=green!35]{}{1e-02} & \Block[fill=green!35]{}{1e-02} & \Block[fill=green!35]{}{2e-02} \\ \hline
\Block[fill=gray!20]{2-1}{\textbf{$\mathbf{nC_7H_{16}}$}} & \Block[fill=Lavender]{}{\textbf{PCA}} & \Block[fill=green!35]{}{2e-04} & \Block[fill=white]{}{5e-04} & \Block[fill=white]{}{4e-04} & \Block[fill=green!35]{}{3e-04} & \Block[fill=green!35]{}{4e-03} & \Block[fill=white]{}{4e-04} & \Block[fill=white]{}{1e-03} & \Block[fill=white]{}{9e-04} & \Block[fill=white]{}{2e-03} & \Block[fill=white]{}{6e-04}\\ 
 & \Block[fill=Apricot!35]{}{\textbf{COK}} & \Block[fill=white]{}{4e-04} & \Block[fill=green!35]{}{3e-04} & \Block[fill=green!35]{}{3e-04} & \Block[fill=white]{}{3e-04} & \Block[fill=white]{}{8e-03} & \Block[fill=green!35]{}{3e-04} & \Block[fill=green!35]{}{6e-04} & \Block[fill=green!35]{}{5e-04} & \Block[fill=green!35]{}{4e-04} & \Block[fill=green!35]{}{4e-04} \\ \hline
\Block[fill=gray!20]{2-1}{\textbf{$\mathbf{OH}$}} & \Block[fill=Lavender]{}{\textbf{PCA}} & \Block[fill=white]{}{3e-02} & \Block[fill=white]{}{2e-02} & \Block[fill=green!35]{}{1e-02} & \Block[fill=white]{}{1e-02} & \Block[fill=green!35]{}{9e-02} & \Block[fill=white]{}{4e-02} & \Block[fill=white]{}{3e-02} & \Block[fill=white]{}{5e-02} & \Block[fill=white]{}{3e-02} & \Block[fill=white]{}{6e-02}\\ 
 & \Block[fill=Apricot!35]{}{\textbf{COK}} & \Block[fill=green!35]{}{2e-02} & \Block[fill=green!35]{}{1e-02} & \Block[fill=white]{}{2e-02} & \Block[fill=green!35]{}{8e-03} & \Block[fill=white]{}{1e-01} & \Block[fill=green!35]{}{2e-02} & \Block[fill=green!35]{}{2e-02} & \Block[fill=green!35]{}{2e-02} & \Block[fill=green!35]{}{2e-02} & \Block[fill=green!35]{}{1e-02} \\ \hline
\Block[fill=gray!20]{2-1}{\textbf{T}} & \Block[fill=Lavender]{}{\textbf{PCA}} & \Block[fill=white]{}{1e-02} & \Block[fill=white]{}{2e-02} & \Block[fill=white]{}{2e-02} & \Block[fill=white]{}{1e-02} & \Block[fill=green!35]{}{4e-02} & \Block[fill=white]{}{2e-02} & \Block[fill=white]{}{1e-02} & \Block[fill=white]{}{2e-02} & \Block[fill=white]{}{1e-02} & \Block[fill=white]{}{2e-02}\\ 
 & \Block[fill=Apricot!35]{}{\textbf{COK}} & \Block[fill=green!35]{}{1e-02} & \Block[fill=green!35]{}{8e-03} & \Block[fill=green!35]{}{1e-02} & \Block[fill=green!35]{}{7e-03} & \Block[fill=white]{}{9e-02} & \Block[fill=green!35]{}{1e-02} & \Block[fill=green!35]{}{1e-02} & \Block[fill=green!35]{}{1e-02} & \Block[fill=green!35]{}{1e-02} & \Block[fill=green!35]{}{1e-02} \\ \hline
\Block[draw=red,line-width=1.5pt]{2-12}{} \Block[fill=gray!20]{2-1}{\textbf{HRR}} & \Block[fill=Lavender]{}{\textbf{PCA}} & \Block[fill=white]{}{7e-02} & \Block[fill=white]{}{6e-02} & \Block[fill=white]{}{5e-02} & \Block[fill=white]{}{8e-02} & \Block[fill=green!35]{}{6e-02} & \Block[fill=white]{}{9e-02} & \Block[fill=white]{}{4e-02} & \Block[fill=white]{}{2e-01} & \Block[fill=white]{}{1e-01} & \Block[fill=white]{}{4e-01}\\ 
 & \Block[fill=Apricot!35]{}{\textbf{COK}} & \Block[fill=green!35]{}{2e-02} & \Block[fill=green!35]{}{4e-02} & \Block[fill=green!35]{}{4e-02} & \Block[fill=green!35]{}{2e-02} & \Block[fill=white]{}{8e-02} & \Block[fill=green!35]{}{5e-02} & \Block[fill=green!35]{}{4e-02} & \Block[fill=green!35]{}{5e-02} & \Block[fill=green!35]{}{3e-02} & \Block[fill=green!35]{}{5e-02} \\ \hline
\end{NiceTabular}
\vspace{0.1cm}
\caption{Cumulative errors ($\epsilon$) in the ignition zone for test data with {\node} solver for \textit{n}heptane-air mixture.}
\label{tab:errors_ignition_heptane}
\end{table*}

\begin{table*}[h!] \scriptsize
\centering

\newcolumntype{?}{!{\vrule width 0pt}}
\begin{NiceTabular}{?c?c?c?c?c?c?c?c?c?c?c?c?}
\hline
\multicolumn{2}{?c?}{\cellcolor{gray!20}\textbf{T}} & \Block[fill=gray!20]{}{\textbf{1129}} & \Block[fill=gray!20]{}{\textbf{1250}} & \Block[fill=gray!20]{}{\textbf{1207}} & \Block[fill=gray!20]{}{\textbf{1129}} & \Block[fill=gray!20]{}{\textbf{1102}} & \Block[fill=gray!20]{}{\textbf{1290}} & \Block[fill=gray!20]{}{\textbf{1190}} & \Block[fill=gray!20]{}{\textbf{1262}} & \Block[fill=gray!20]{}{\textbf{1150}} & \Block[fill=gray!20]{}{\textbf{1300}} \\ \hline
\multicolumn{2}{?c?}{\cellcolor{gray!20}$\boldsymbol{\phi}$} & \Block[fill=gray!20]{}{\textbf{0.56}} & \Block[fill=gray!20]{}{\textbf{0.58}} & \Block[fill=gray!20]{}{\textbf{0.59}} & \Block[fill=gray!20]{}{\textbf{0.64}} & \Block[fill=gray!20]{}{\textbf{0.6}} & \Block[fill=gray!20]{}{\textbf{0.6}} & \Block[fill=gray!20]{}{\textbf{0.78}} & \Block[fill=gray!20]{}{\textbf{0.82}} & \Block[fill=gray!20]{}{\textbf{0.8}} & \Block[fill=gray!20]{}{\textbf{0.8}} \\ \hline
\Block[fill=gray!20]{2-1}{\textbf{$\mathbf{H_2O}$}} & \Block[fill=Lavender]{}{\textbf{PCA}} & \Block[fill=white]{}{4e-02} & \Block[fill=green!35]{}{3e-02} & \Block[fill=green!35]{}{2e-02} & \Block[fill=white]{}{4e-02} & \Block[fill=green!35]{}{1e-01} & \Block[fill=white]{}{4e-02} & \Block[fill=green!35]{}{1e-02} & \Block[fill=green!35]{}{9e-03} & \Block[fill=green!35]{}{1e-02} & \Block[fill=white]{}{3e-02}\\ 
 & \Block[fill=Apricot!35]{}{\textbf{COK}} & \Block[fill=green!35]{}{4e-02} & \Block[fill=white]{}{3e-02} & \Block[fill=white]{}{2e-02} & \Block[fill=green!35]{}{4e-02} & \Block[fill=white]{}{3e-01} & \Block[fill=green!35]{}{1e-02} & \Block[fill=white]{}{2e-02} & \Block[fill=white]{}{9e-03} & \Block[fill=white]{}{3e-02} & \Block[fill=green!35]{}{9e-03} \\ \hline
\Block[fill=gray!20]{2-1}{\textbf{$\mathbf{nC_7H_{16}}$}} & \Block[fill=Lavender]{}{\textbf{PCA}} & \Block[fill=green!35]{}{7e-03} & \Block[fill=green!35]{}{4e-03} & \Block[fill=green!35]{}{2e-03} & \Block[fill=green!35]{}{7e-03} & \Block[fill=green!35]{}{9e-03} & \Block[fill=green!35]{}{5e-03} & \Block[fill=white]{}{8e-03} & \Block[fill=white]{}{6e-03} & \Block[fill=green!35]{}{4e-03} & \Block[fill=green!35]{}{5e-03}\\ 
 & \Block[fill=Apricot!35]{}{\textbf{COK}} & \Block[fill=white]{}{1e-02} & \Block[fill=white]{}{5e-03} & \Block[fill=white]{}{7e-03} & \Block[fill=white]{}{1e-02} & \Block[fill=white]{}{5e-02} & \Block[fill=white]{}{7e-03} & \Block[fill=green!35]{}{6e-03} & \Block[fill=green!35]{}{5e-03} & \Block[fill=white]{}{5e-03} & \Block[fill=white]{}{5e-03} \\ \hline
\Block[fill=gray!20]{2-1}{\textbf{$\mathbf{OH}$}} & \Block[fill=Lavender]{}{\textbf{PCA}} & \Block[fill=green!35]{}{2e-02} & \Block[fill=green!35]{}{9e-03} & \Block[fill=green!35]{}{5e-03} & \Block[fill=green!35]{}{1e-02} & \Block[fill=green!35]{}{2e-02} & \Block[fill=green!35]{}{1e-02} & \Block[fill=white]{}{2e-02} & \Block[fill=green!35]{}{1e-02} & \Block[fill=white]{}{2e-02} & \Block[fill=white]{}{2e-02}\\ 
 & \Block[fill=Apricot!35]{}{\textbf{COK}} & \Block[fill=white]{}{2e-02} & \Block[fill=white]{}{3e-02} & \Block[fill=white]{}{4e-02} & \Block[fill=white]{}{2e-02} & \Block[fill=white]{}{2e-01} & \Block[fill=white]{}{2e-02} & \Block[fill=green!35]{}{1e-02} & \Block[fill=white]{}{2e-02} & \Block[fill=green!35]{}{1e-02} & \Block[fill=green!35]{}{6e-03} \\ \hline
\Block[fill=gray!20]{2-1}{\textbf{T}} & \Block[fill=Lavender]{}{\textbf{PCA}} & \Block[fill=white]{}{2e-02} & \Block[fill=white]{}{8e-03} & \Block[fill=white]{}{7e-03} & \Block[fill=white]{}{2e-02} & \Block[fill=green!35]{}{7e-02} & \Block[fill=white]{}{1e-02} & \Block[fill=white]{}{9e-03} & \Block[fill=green!35]{}{6e-03} & \Block[fill=green!35]{}{8e-03} & \Block[fill=white]{}{1e-02}\\ 
 & \Block[fill=Apricot!35]{}{\textbf{COK}} & \Block[fill=green!35]{}{8e-03} & \Block[fill=green!35]{}{3e-03} & \Block[fill=green!35]{}{4e-03} & \Block[fill=green!35]{}{2e-02} & \Block[fill=white]{}{1e-01} & \Block[fill=green!35]{}{9e-03} & \Block[fill=green!35]{}{8e-03} & \Block[fill=white]{}{7e-03} & \Block[fill=white]{}{8e-03} & \Block[fill=green!35]{}{8e-03} \\ \hline
\Block[draw=red,line-width=1.5pt]{2-12}{} \Block[fill=gray!20]{2-1}{\textbf{HRR}} & \Block[fill=Lavender]{}{\textbf{PCA}} & \Block[fill=white]{}{1e-02} & \Block[fill=white]{}{2e-02} & \Block[fill=white]{}{2e-02} & \Block[fill=white]{}{2e-02} & \Block[fill=green!35]{}{2e-02} & \Block[fill=white]{}{4e-02} & \Block[fill=green!35]{}{1e-02} & \Block[fill=white]{}{5e-02} & \Block[fill=green!35]{}{1e-02} & \Block[fill=white]{}{1e-01}\\ 
 & \Block[fill=Apricot!35]{}{\textbf{COK}} & \Block[fill=green!35]{}{9e-03} & \Block[fill=green!35]{}{2e-02} & \Block[fill=green!35]{}{2e-02} & \Block[fill=green!35]{}{1e-02} & \Block[fill=white]{}{6e-02} & \Block[fill=green!35]{}{2e-02} & \Block[fill=white]{}{1e-02} & \Block[fill=green!35]{}{2e-02} & \Block[fill=white]{}{2e-02} & \Block[fill=green!35]{}{3e-02} \\ \hline
\end{NiceTabular}
\vspace{0.1cm}
\caption{Cumulative errors ($\epsilon$) in the entire time interval for test data with {\node} solver for \textit{n}heptane-air mixture.}
\label{tab:errors_entire_heptane}
\end{table*}

To demonstrate the robustness of {\cokpca} across different fuels, we now consider autoignition of \textit{n}heptane-air mixture. The chemistry in the thermochemical space is represented with 88 species and 387 reactions mechanism \cite{yoo2011direct}. The dataset encompasses 33 distinct initial conditions for training, 10 randomly selected initial conditions for validation, and the 10 configurations farthest from the training and validation data for final model assessment. {\pca} and {\cokpca} based manifolds are constructed using thermochemical scalar data from nine of the 33 training configurations. The covariance matrix and cokurtosis tensor are generated using Pareto-scaled data. After obtaining the principal components (PCs), the dimensionality of the LDMs is set to $n_q=10$. Only the {\node} solver is employed to integrate the PCs from the initial conditions.

Figure~\ref{YTs node heptane} illustrates the temporal profiles of the reconstructed mass fractions of the selected species, temperature, and computed heat release rate (HRR) from thermochemical scalars. Unlike the ethylene-air case, {\pca} exhibits a greater deviation from the actual trajectories in this case, particularly spurious spikes appear in the ignition zone for species and temperature. On the other hand, {\cokpca} provides a good agreement with the actual trajectories. In the HRR profiles, {\pca} significantly overestimated the peak value and exhibits spurious endothermic behavior. These observations are further substantiated by the cumulative errors presented in Tabs.~\ref{tab:errors_ignition_heptane} and~\ref{tab:errors_entire_heptane}, which reveals that {\cokpca} generally captures the HRR more accurately than {\pca}. These findings support the hypothesis that {\cokpca} captures chemical dynamics in the ignition zone more effectively than {\pca}.

\section{Conclusions\label{sec:conclusions}} \addvspace{10pt}

This study compares the performance of cokurtosis principal component analysis (CoK-PCA) and standard principal component analysis (PCA) for dimensionality reduction in an a posteriori setting. Two test cases that capture spontaneous ignition in a homogeneous reactor were examined to investigate the efficacy of low-dimensional manifolds. The governing ordinary differential equations (ODEs) in reduced space were solved using a standard ODE solver and neural ODE solver, where the source terms were modeled using artificial neural networks (ANN). A posteriori analysis demonstrated the effectiveness of CoK-PCA based low-dimensional manifolds in accurately representing the ignition processes. The neural ODE solver, which incorporates time integration into ANN training, outperformed the standard ODE solver by minimizing propagation errors and providing more accurate results. These findings underscore the potential of CoK-PCA as a valuable tool for modeling complex reacting flow systems. In future research, we aim to extend this study by incorporating the transport phenomenon, which includes diffusive and advective contributions in addition to the reaction dynamics, and assess the efficacy of CoK-PCA.

\section*{Acknowledgements}
The work done at IISc is supported by ANRF Core Research Grant.



\bibliographystyle{main-style}
\bibliography{main}





\end{document}